\documentclass[12pt]{iopart}

\usepackage{iopams} 

\usepackage[colorlinks, pdfborder={0 0 0}, plainpages=false,linkcolor=blue,citecolor=blue,urlcolor=magenta]{hyperref}
\usepackage{graphicx}
\usepackage[usenames,dvipsnames]{color}
\usepackage[normalem]{ulem} 

\begin{document}

\newcommand{\red}{\textcolor{red}}
\newcommand{\dan}[1]{\textcolor{WildStrawberry}{#1}}
\newcommand{\Mark}[1]{\textcolor{Cerulean}{#1}}
\newcommand{\larry}[1]{\textcolor{OliveGreen}{#1}}
\newcommand{\bela}[1]{\textcolor{Blue}{#1}}
\newcommand{\rana}[1]{\textcolor{RedOrange}{#1}}
\newcommand{\nick}[1]{\textcolor{Purple}{#1}}

\newcommand{\Note}[1]{\textcolor{blue}{\textbf{[#1]}}}

\newcommand{\Caltech}{\address{Theoretical Astrophysics 350-17,
  California Institute of Technology, Pasadena, CA 91125, USA}}
\newcommand{\Cornell}{\address{Center for Radiophysics and Space
  Research, Cornell University, Ithaca, New York 14853, USA}}
\newcommand{\CITA}{\address{Canadian Institute for Theoretical
  Astrophysics, 60 St.~George Street, University of Toronto,
  Toronto, ON M5S 3H8, Canada}} %
\newcommand{\GWPAC}{\address{Gravitational Wave Physics and
  Astronomy Center, California State University Fullerton,
  Fullerton, California 92834, USA}} %

\title[Numerical Brownian thermal noise in amorphous and crystalline coatings]{Numerically modeling Brownian thermal noise in amorphous and crystalline thin coatings}

\author{Geoffrey Lovelace} \GWPAC \ead{glovelace@fullerton.edu}
\author{Nicholas Demos} \GWPAC \ead{ndemos@csu.fullerton.edu}
\author{Haroon Khan} \GWPAC \ead{haroon.kh@csu.fullerton.edu}

\begin{abstract}
Thermal noise is expected to be one of the noise sources limiting the astrophysical reach of Advanced LIGO (once commissioning is complete) and third-generation detectors. Adopting crystalline materials for thin, reflecting mirror coatings, rather than the amorphous coatings used in current-generation detectors, could potentially reduce thermal noise. Understanding and reducing thermal noise requires accurate theoretical models, but modeling thermal noise analytically is especially challenging with crystalline materials. Thermal noise models typically rely on the fluctuation-dissipation theorem, which relates the power spectral density of the thermal noise to an auxiliary elastic problem. 
In this paper, we present results from a new, open-source tool that numerically solves the auxiliary elastic problem to compute the Brownian thermal noise for both amorphous and crystalline coatings. We employ the open-source deal.ii and PETSc frameworks to solve the auxiliary elastic problem using a finite-element method, adaptive mesh refinement, and parallel processing that enables us to use high resolutions capable of resolving the thin reflective coating. We verify numerical convergence, and by running on up to hundreds of compute cores, we resolve the coating elastic energy in the auxiliary problem to approximately 0.1\%. We compare with approximate analytic solutions for amorphous materials, and we verify that our solutions scale as expected with changing beam size, mirror dimensions, and coating thickness. Finally, we model the crystalline coating thermal noise in an experiment reported by Cole and collaborators (2013), comparing our results to a simpler numerical calculation that treats the coating as an ``effectively amorphous'' material. We find that treating the coating as a cubic crystal instead of as an effectively amorphous material increases the thermal noise by about 3\%. Our results are a step toward better understanding and reducing thermal noise to increase the reach of future gravitational-wave detectors.
\end{abstract}

\maketitle

\section{Introduction}
In 2015, the Advanced Laser Interferometer Gravitational-Wave Observatory (Advanced LIGO) detected gravitational waves passing through Earth for the first time, inaugurating the era of gravitational-wave astronomy~\cite{Abbott:2016blz,Abbott:2016nmj,TheLIGOScientific:2016pea}. During 2017, Advanced LIGO began its second observation run; during this run, LIGO observed another gravitational wave from merging black holes~\cite{scientific2017gw170104}. Other second-generation gravitational-wave observatories, such as Advanced Virgo~\cite{TheVirgo:2014hva}, the Kamioka Gravitational-wave detector (KAGRA)~\cite{Somiya:2011np,aso2013interferometer}, and LIGO India~\cite{LIGOIndia}, will soon join Advanced LIGO, helping to better constrain the sky location and the properties
of observed gravitational waves' sources. Third-generation detector designs, such as the
Einstein Telescope~\cite{EinsteinTelescope} and Cosmic Explorer~\cite{abbott2017exploring}, aim to gain a factor of
$\approx 10$ in sensitivity over second-generation detectors, which corresponds to a factor of $1000$ in detection rate. 

\begin{figure*}[ht]
\begin{center}
\includegraphics[width=4.5in]{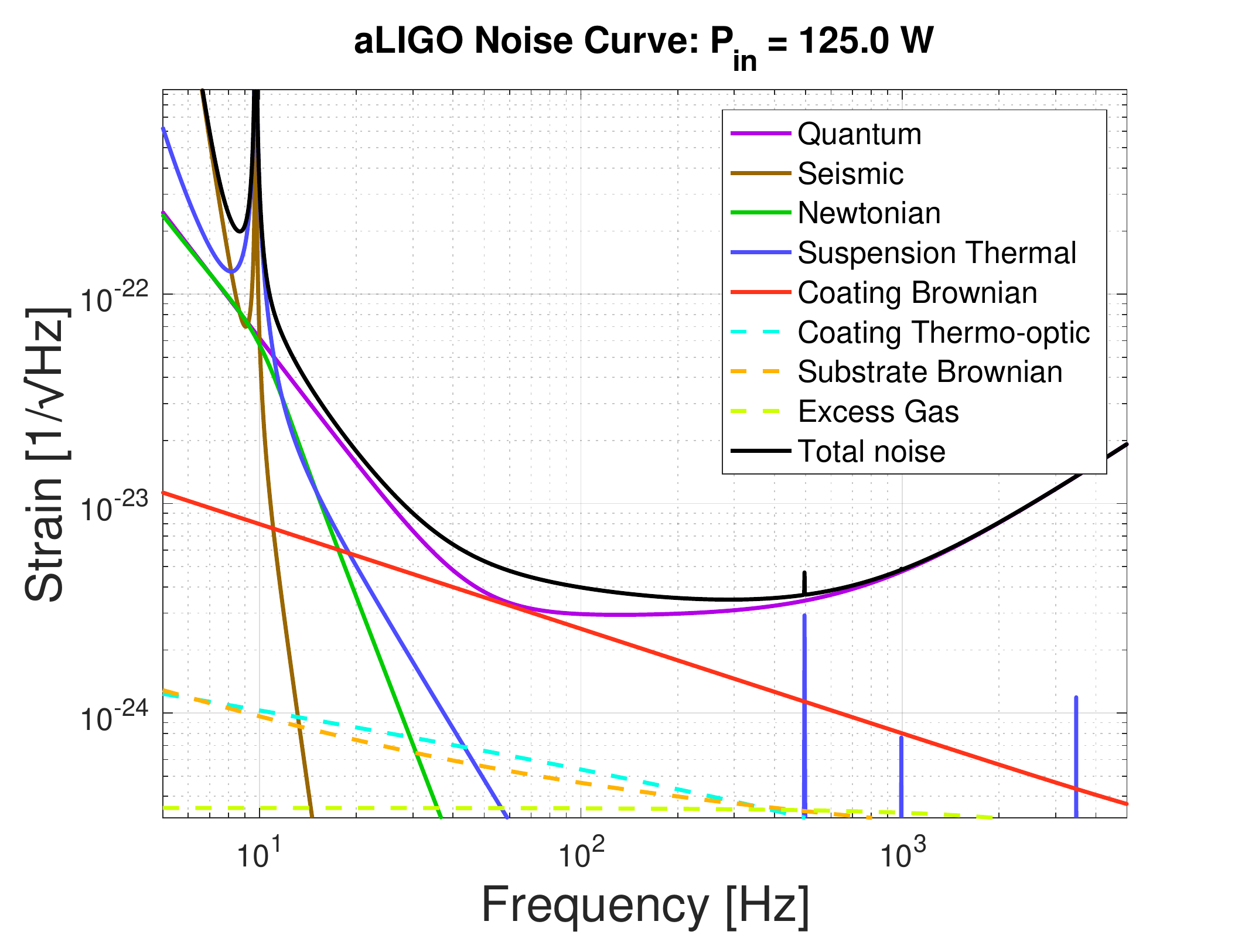}
\end{center}
\caption{
Principal noise terms for Advanced LIGO as seen in ~Fig. 2 of Ref.~\cite{aLIGO2}, updated as of November 16, 2016, using the Gravitational Wave Interferometer Noise Calculator~\cite{gwinc}. The component noises add as a root-square-sum since they are statistically independent.\label{fig:gwinc}}
\end{figure*}

Thermal noise is expected to be one of the noise sources that limits the sensitivity of second-generation detectors (once Advanced LIGO commissioning is complete) and third-generation ground-based gravitational-wave detectors. The total thermal noise budget includes contributions from i) Brownian thermal noise, caused by mechanical losses (i.e., by small, imaginary, dissipative terms of the material's elastic moduli), and ii) thermoelastic and thermorefractive coating noise, caused by temperature fluctuations in the materials. Figure~\ref{fig:gwinc} shows an Advanced LIGO noise curve, computed from the constituent noises in the instrument, current as of November 2016. The figure shows that Brownian coating thermal noise and quantum noise are the most important noises in Advanced LIGO's most sensitive frequency band ($\sim 100\mbox{ Hz}$).

Brownian coating thermal noise 
limits the sensitivity of Advanced LIGO (e.g.~Fig. 2 of Ref.~\cite{aLIGO2}) 
and third generation detectors (e.g.~Fig.~20 of 
Ref.~\cite{adhikari2014gravitational}). Therefore, a substantial research effort is studying Brownian coating thermal noise 
theoretically and experimentally~\cite{harry2002thermal,
penn2003mechanical,rowan2005thermal,harry2006thermal,harry2007titania,
flaminio2010study,kondratiev2011thermal,evans2012reduced,hong2013brownian,
bassiri2013correlations,murray2017cryogenic,gras2017audio,kroker2017brownian}.
Thermoelastic and 
thermorefractive noise (caused by temperature fluctuations) in the 
coating~\cite{braginsky2003thermodynamical,fejer2004thermoelastic,
evans2008thermo} can also significantly contribute to the total noise in future gravitational-wave detectors, depending on the optics' materials and temperatures, as can Brownian and thermoelastic noise in the 
substrate~\cite{braginsky2000thermo,
liu2000thermoelastic,o2004implications,vinet2005mirror} 
and suspensions~\cite{saulson1990thermal,gillespie1993thermal,
braginsky1999reduce,gonzalez2000suspensions,kumar2008finite}.
Multi-layered coatings can be designed so that 
thermoelastic and thermorefractive noise largely 
cancel~\cite{evans2008thermo,Harry:2011book},
since these noises add coherently. Photothermal noise (temperature 
fluctuations in the coating caused by absorption of incident laser power) 
can similarly be coherently canceled~\cite{chalermsongsak2015coherent}. 

Even though our primary motivation for modeling thermal noise is the application to gravitational-wave detection, we note that thermal noise is also a limiting noise in a number of other applications. For instance, thermal noise is a limiting noise for atomic clocks
(e.g.~Ref.~\cite{jiang2011making}). See, e.g., Ref.~\cite{harry2012optical} for a broader introduction to thermal noise.

Theoretical predictions are essential tools for understanding and minimizing thermal noise. Thermal-noise models typically rely on 
the fluctuation-dissipation 
theorem~\cite{callen1951irreversibility,bernard1959fdt,kubo1966fluctuation}, 
which relates the thermal 
noise to the solution of an auxiliary elastic 
problem~\cite{gonzalez1994brownian,levin1998internal}. 
When a sinusoidally varying pressure with the same spatial distribution 
as the laser beam 
shape (i.e.~intensity profile) is applied, power is dissipated as the mirror elastically 
deforms. The fluctuation-dissipation theorem relates the dissipated power in the auxiliary elastic problem to the spectral density of the mirror's thermal fluctuations. 
In ground-based gravitational-wave detectors, a sound wave crosses 
the mirror in much less time 
than a gravitational-wave period; 
therefore, thermal noise models often make 
a quasistatic approximation~\cite{levin1998internal}, 
with the dissipated energy per cycle 
becoming a product 
of the potential energy of the elastostatic deformation and a 
mechanical loss angle. 

Existing models of thermal noise almost always treat the elastic properties 
of the materials isotropically when applying the fluctuation-dissipation theorem; this allows the elastic problem 
to be approached analytically. This is perfectly sensible for amorphous 
materials (as used in Advanced LIGO~\cite{aLIGO2}), but  
most plans for future ground-based detectors use crystalline 
mirror substrates~\cite{hirose2014update,punturo2010einstein,
adhikari2014gravitational}. 
KAGRA~\cite{Somiya:2011np}, 
which aims to reduce thermal noise by lowering the detector 
temperature to 20 K, will use crystalline sapphire 
substrates, motivated by crystalline sapphire's high thermal 
conductivity~\cite{Somiya:2011np} 
and fused silica's increased mechanical loss at low 
temperatures~\cite{schroeter2007mechanical}. 
Also, GaAs:Al$_{0.92}$Ga$_{0.08}$As (AlGaAs) 
crystalline coatings experimentally show 
great promise for reducing Brownian coating thermal 
noise~\cite{cole2013tenfold,cole2016highperformance}.

Developing a theoretical model of Brownian 
coating mirror thermal noise, one flexible enough to incorporate both 
amorphous and crystalline materials, will help to understand and reduce Brownian coating noise and 
thus to extend the reach of future ground-based 
gravitational-wave detectors. To correctly understand thermal noise in crystalline materials, these models must account for their anisotropic elastic moduli. However, crystalline 
materials' elastic properties are anisotropic, making analytic calculation of 
elastic deformation in crystalline materials highly nontrivial 
(though recent work~\cite{pang2009effect} 
has succeeded in yielding a semi-analytic 
solution for the static elastic deformation of a semi-infinite cubic 
crystal). The formidable challenges toward an analytic model of 
crystal coating thermal noise motivate numerical thermal-noise modeling. A recent study has numerically modeled 
thermal noise in crystalline substrates~\cite{heinert2014fluctuation}.

In this paper, we numerically calculate Brownian coating and substrate noise. We present a new, open-source tool, based on existing open-source  frameworks, that i) solves the auxiliary elastic problem for a cylindrical mirror with a thin coating and ii) uses the solution and the fluctuation-dissipation theorem to compute the power spectral density of the thermal noise. We adopt a finite-element method, a typical approach in elasticity computations, and we use adaptive mesh refinement and parallel processing to resolve the thin coating. 

Specifically, we compute the thermal noise on a cylindrical  mirror with a single-layer, thin reflective coating. For concreteness, we focus on the same case (i.e., same mirror dimensions, coating thickness, laser beam width, temperature, and elastic properties) considered by Cole and collaborators in Ref.~\cite{cole2013tenfold}, to facilitate comparison with their previous calculations (which assumed isotropic materials).

In Sec.~\ref{sec:performance}, we show how our code's run time scales with the number of compute cores. We find that running on 50-100 cores greatly improved performance, but further increases do not significantly improve performance (perhaps because of increasing communication costs). We also find that running on hundreds of cores enabled us to reach higher resolutions, by increasing the total memory available for the calculation.

In Sec.~\ref{sec:results}, we first test our code's numerical convergence. We find that the elastic internal energy (obtained from our solution of the elastic displacement vector) converges with increasing resolution, and we estimate our numerical uncertainty in the coating energy is 0.1\%. Then, we compare our code's results for amorphous materials to approximate, analytic solutions for the amorphous case. We find that edge effects and coating thickness effects scale as we expect. 

Then, we compute the Brownian coating thermal noise for i) a mirror with an amorphous, fused-silica substrate and a crystalline, AlGaAs coating, and ii) a mirror with the same substrate but with the crystalline coating replaced with an ``effective isotropic'' coating, i.e., with an amorphous coating with elastic properties meant to mimic the AlGaAs's crystalline elastic properties. The thermal noise, treating the coating as a cubic crystal, is approximately 3\% larger than the thermal noise when treating the coating as an effective isotropic material. 

We also compare our numerical calculations with an approximate, analytic result for a semi-infinite, amorphous mirror with a thin, amorphous coating. The effective isotropic calculation predicts approximately 7\% smaller thermal noise, because of finite-size effects. The cubic-crystal, numerical, coating thermal noise differs from the approximate solution by approximately 4\%. From these results, we conclude that, for our calculation, neglecting crystal effects introduces an error of the same order as neglecting edge effects.

The rest of this paper is organized as follows. In Sec.~\ref{sec:techniques}, we introduce our notation and detail our numerical methods for computing the Brownian substrate and coating thermal noise for a mirror with a single-layer, thin, reflective, possibly crystalline coating. In Sec.~\ref{sec:results}, we present our physical results after presenting results that verify our tool's performance and scaling. Finally, we briefly conclude in Sec.~\ref{sec:conclusion}.

\section{Techniques}
\label{sec:techniques}
\subsection{Mirror geometry and laser beam intensity profile}

We begin by considering a cylindrical mirror of radius $R$ and height $H$, where the mirror dimensions are comparable: $R\sim H$. A thin, reflective coating of thickness $d$ satisfying $d \ll R$ and $d \ll H$ covers the front face of the mirror. (In practice, LIGO mirror coatings consist of even thinner alternating layers of different materials; here, for simplicity we only consider single-layer coatings, leaving multi-layer coatings for future work.) We typically will choose our coordinates so that the $z$ axis is the axis of symmetry of the cylinder, where the coating-substrate interface lies in the plane $z=0$. 

LIGO measures the position of the mirror by shining a laser beam on the mirror's surface; the beam measures $q(t)$, the surface position weighted by the beam's intensity:
\begin{eqnarray}
q(t) \equiv \int_0^{2\pi} d\varphi \int_0^R dr r p(r,\varphi) Z(r,\varphi,t).
\end{eqnarray} Here, $Z(r,\varphi,t)$ is the displacement at time $t$ of a point on the mirror's surface at cylindrical coordinates $(r,\varphi)$, in the direction parallel to the incident laser beam (which we take to travel along the $z$ axis). 

Typically, we will center the beam profile $p(r,\varphi)$ on the mirror, so that the beam's intensity profile is 
\begin{eqnarray}\label{eq:pressureGauss}
p(r) & = & \frac{1}{\pi r_0^2\left(1-e^{-R^2/r_0^2}\right)} e^{-r^2/r_0^2}, 
\end{eqnarray} normalized so that
\begin{eqnarray}
\oint dA p(r) = 2 \pi \int_0^R dr r p(r) = 1.
\end{eqnarray} In LIGO, to minimize diffraction losses, the beam size $r_0$ is kept significantly smaller than the mirror radius: $r_0 \ll R$. Therefore, in practice we neglect the Gaussian in the denominator of Eq.~(\ref{eq:pressureGauss}) when normalizing $p(r)$. Unless stated otherwise, in the rest of this paper, we choose an intensity profile
\begin{eqnarray} 
p(r) & = & \frac{1}{\pi r_0^2} e^{-r^2/r_0^2},
\end{eqnarray} Note that some references, such as Ref.~\cite{harry2002thermal} and Ref.~\cite{cole2013tenfold}, use a beam width $w=\sqrt{2} r_0$. In terms of $w$, the intensity profile becomes
\begin{eqnarray}
p(r) & = & \frac{2}{\pi w^2} e^{-2 r^2/w^2}.
\end{eqnarray}

\subsection{Fluctuation-dissipation theorem}
Because the mirror has a temperature $T$, internal thermal noise in the mirror causes fluctuations in $Z(r,\varphi,t)$. We compute the single-sided power\footnote{Note that occasionally we will refer to the amplitude spectral density, which is the square root of the power spectral density.} spectral density $S_q(f)$ of the thermal noise associated with the measurement $q$ at frequency $f$ using the fluctuation-dissipation theorem~\cite{callen1951irreversibility,bernard1959fdt,kubo1966fluctuation,gonzalez1994brownian,levin1998internal}. The theorem relates $S_q(f)$ to the solution of an auxiliary elastic problem:
\begin{enumerate}
\item Imagine applying a pressure to the mirror face (i.e., the top of the mirror coating) by pushing on it with a force $F p(r,\varphi) \sin 2\pi f t$, where $F$ is a force amplitude and $p(r,\varphi)$ is the same as the intensity profile of the laser beam in the actual measurement.
\item The mirror deforms, dissipating energy at a rate\footnote{Note that in Ref.~\cite{hong2013brownian}, $W_{\rm diss}$ refers not to the dissipated power but to the energy dissipated in each cycle. In this paper, $W_{\rm diss}$ refers to dissipated power averaged over a cycle, and $E_{\rm diss}$ refers to the energy dissipated in one cycle.}
(averaged over a cycle) $W_{\rm diss}$.
\item The fluctuation-dissipation theorem gives the thermal noise $S_q(f)$ associated with the actual measurement $q$ in terms of $W_{\rm diss}$ calculated in the auxiliary elastic problem:
\begin{eqnarray}
S_q(f) & = & \frac{2 k_B T}{\pi^2 f^2} \frac{W_{\rm diss}}{F^2}.
\end{eqnarray} Here $k_B$ is Boltzmann's constant. Note that $W_{\rm diss}\propto F^2$, so the thermal noise does not depend on $F$. 
\end{enumerate} 

\subsection{Elasticity}

For applications to LIGO, the thermal noise is relevant at frequencies $f\sim 100\mbox{ Hz}$ where LIGO is most sensitive. Because these frequencies are far below the resonant frequencies $f\sim 10^4\mbox{ Hz}$ of the mirror materials, the applied force can be treated quasistatically. In the quasistatic approximation, the oscillating applied force is replaced with a static force. 

Applying this static force to the face of the mirror deforms the mirror. A small element in the mirror at position $x_i$ is displaced by $u_i$:
\begin{eqnarray}
 x_i \rightarrow x_i + u_i.
\end{eqnarray}
This leads to a strain 
\begin{eqnarray}\label{eq:strain}
S_{ij} & = & \nabla_{(i}u_{j)} \equiv \frac{1}{2}\left(\nabla_i u_j + \nabla_j u_i \right).
\end{eqnarray} Here and throughout this paper, indices $i,j,k,\ldots$ are spatial indices running over Cartesian coordinates $x,y,z$. By choosing $F$ to be sufficiently small, the deformation can be made sufficiently small that the material's deformation is within its elastic limit. That is, the applied stress is proportional to the strain (``Hooke's law''): 
\begin{eqnarray}\label{eq:Hooke}
T_{ij} = Y_{ijkl} S_{kl},
\end{eqnarray} where here and throughout this paper we adopt the Einstein summation convention, summing over repeated indices. The Young's tensor $Y_{ijkl}$ is symmetric on each pair of indices and on interchanging the pairs of indices,
\begin{eqnarray}
Y_{ijkl} & = & Y_{jikl} = Y_{ijlk} = Y_{klij},
\end{eqnarray} leaving 21 independent components of $Y_{ijkl}$. In this paper, although our tool and methods are implemented for any $Y_{ijkl}$, we primarily confine our attention to two cases:
\begin{enumerate}
\item amorphous materials, where symmetry leaves only 2 independent components in $Y_{ijkl}$, corresponding to the Young's modulus $Y$ and Poisson ratio $\sigma$, and 
\item cubic crystalline materials, which have 3 independent components in $Y_{ijkl}$.
\end{enumerate} In both cases, the nonzero Young's tensor components can be written in terms of three elastic moduli
$c_{11}$, $c_{12}$, and $c_{44}$ as follows:
\begin{eqnarray}
c_{11} & = & Y_{xxxx} = Y_{yyyy} = Y_{zzzz},\\
c_{12} & = & Y_{xxyy} = Y_{xxzz} = Y_{yyxx} = Y_{yyzz} = Y_{zzxx} = Y_{zzyy},\\
c_{44} & = & Y_{xyxy} = Y_{xyyx} = Y_{yxxy} = Y_{yxyx} = Y_{xzxz} = Y_{xzzx}\nonumber\\ & = & Y_{zxxz} = Y_{zxzx} = Y_{yzyz} = Y_{zyyz} = Y_{zyzy} = Y_{yzzy}.
\end{eqnarray} For cubic crystals, $c_{11}$, $c_{12}$, and $c_{44}$ are independent, while for amorphous materials, 
\begin{eqnarray}
c_{11} & = & \frac{Y(1-\sigma)}{(1+\sigma)(1-2\sigma)},\\
c_{12} & = & \frac{Y \sigma}{(1+\sigma)(1-2\sigma)},\\
c_{44} & = & \frac{Y}{2(1+\sigma)}.
\end{eqnarray}

The stress and strain combine to form the stored potential energy density, such that the total stored potential energy in a volume $\mathcal{V}$ is
\begin{eqnarray}\label{eq:U}
U=-\frac{1}{2}\int_{\mathcal{V}} dV S_{ij} T_{ij}. 
\end{eqnarray}
The dissipated power is then (e.g., inserting inserting Eq.~\ref{eq:U} into Eq.~(12) of Ref.~\cite{levin1998internal})
\begin{eqnarray}
W_{\rm diss} = 2 \pi f \phi(f) U = - \pi f \int dV S_{ij} T_{ij},
\end{eqnarray} where $\phi(f)$ is a (potentially frequency-dependent) loss angle determined by the imaginary, dissipative elastic moduli. Then the thermal noise becomes (Eq.~(46) of Ref.~\cite{hong2013brownian})
\begin{eqnarray}
S_q(f) & = & \frac{4 k_B T}{\pi f} \frac{U \phi(f)}{F^2}
\end{eqnarray} 

For a mirror consisting of a thin reflective coating on top of a substrate, the stored energy is the sum of the energy stored in the substrate and in the coating: $U=U_{\rm sub}+U_{\rm coat}$. When the coating and substrate are different materials, the substrate and coating have different loss angles, $\phi_{\rm sub}$ and $\phi_{\rm coat}$. Then, the thermal noise becomes
\begin{eqnarray}\label{eq:FDT}
S_q(f) & = & \frac{4 k_B T}{\pi f} \frac{U_{\rm sub} \phi_{\rm sub} + U_{\rm coat} \phi_{\rm coat}}{F^2}, 
\end{eqnarray} where the stored energies $U_{\rm sub}$ and $U_{\rm coat}$ are volume integrals over the substrate and coating, respectively:
\begin{eqnarray}
U_{\rm sub} & = & U=-\frac{1}{2}\int_{\mathcal{\rm sub}} dV S_{ij} T_{ij},\label{eq:Usub}\\
U_{\rm coat} & = & U=-\frac{1}{2}\int_{\mathcal{\rm coat}} dV S_{ij} T_{ij}.\label{eq:Ucoat}
\end{eqnarray}

\label{sec:lossangles}
In fact, as in Ref.~\cite{hong2013brownian}, an amorphous coating has two loss angles, $\phi_B$ and $\phi_S$, corresponding to the imaginary parts of the coating's bulk and shear modulus. If $U_{B}$ and $U_{S}$ are the elastic energy corresponding to the bulk and shear portions of the elastic energy, then Eq.~(\ref{eq:FDT}) becomes (cf. Eq.~(57) of Ref.~\cite{hong2013brownian})
\begin{eqnarray}\label{eq:multilosssub}
S_q(f) & = & \frac{4 k_B T}{\pi f} \frac{U_{\rm sub} \phi_{\rm sub} + U_{\rm B} \phi_{\rm B} + U_{\rm S} \phi_{\rm S}}{F^2}.
\end{eqnarray}
A crystal coating has, in principle, different loss angles for each independent component of the Young's tensor $Y_{ijkl}$. For instance, for a cubic or zincblende crystal, one might write
\begin{eqnarray}\label{eq:multilosscoat}
S_q(f) & = & \frac{4 k_B T}{\pi f} \frac{U_{\rm sub} \phi_{\rm sub} + U_{\rm 11} \phi_{\rm 11} + U_{\rm 12} \phi_{\rm 12} + U_{\rm 44} \phi_{\rm 44}}{F^2},
\end{eqnarray} where $\phi_{\rm 11}$, $\phi_{\rm 12}$, and $\phi_{\rm 44}$ are small, imaginary parts of $c_{11}$, $c_{12}$, and $c_{44}$, respectively, while $U_{\rm 11}$, $U_{\rm 12}$, and $U_{44}$ are the portions of the elastic energy corresponding to each elastic modulus.
Abernathy and collaborators have recently made the first measurements of separate bulk and coating loss angles for a material~\cite{abernathyMultipleLossAngles}. While perhaps a generalization of their method will be able to successfully measure the three (or more) loss angles in a crystalline material, this has not yet been done. Therefore, in this paper, we characterize the coating by a single effective loss angle $\phi_{\rm coat}$, determined from experiment, though we look forward to generalizing our results along the lines of Eq.~(\ref{eq:multilosscoat}) once experimental measurements of 3 independent loss angles for crystalline materials are available.

Thus to compute the thermal noise, we first numerically compute the displacement $u_i$ of the mirror given an applied pressure exerted at the top of the coating; this amounts to numerically solving Newton's second law for elastostatics,
\begin{eqnarray}
0 & = & -\nabla_i T_{ij} - f_i.\label{eq:secondLaw}
\end{eqnarray} Because we are seeking a solution where the mirror is in elastostatic equilibrium, we set $f_i=0$ except on the mirror boundary with the applied pressure.

After obtaining the numerical solution for $u_i$, we i) numerically compute its gradient to obtain the strain $S_{ij}$, ii) use Eq.~(\ref{eq:Hooke}) to compute the stress, iii) insert the stress and strain into Eqs.~(\ref{eq:Usub}) and (\ref{eq:Ucoat}) and integrate to compute $U_{\rm sub}$ and $U_{\rm coat}$, and finally iv) insert $U_{\rm sub}$ and $U_{\rm coat}$ into Eq.~(\ref{eq:FDT}) to compute the Brownian thermal noise $S_{q}(f)$.

In the rest of this section, we derive the equations that determine $u_i$ and cast them in a form suitable for numerical solution using finite elements. 

\subsection{Weak form of the elastostatic equations for finite-element numerical solutions}

For completeness, we present the weak form of the three-dimensional elasticity equations that we implemented and used in this paper. The following derivation is not new; it generally follows the derivation given in Sec.~2.4.3 of Ref.~\cite{ibrahimbegovic2009nonlinear} except that we prefer different notation.

Consider an applied force deforming a mirror occupying a volume $V$
enclosed by a surface boundary $\Gamma$. 

Note that while eventually we will choose $f_i = 0$ except on the part of $\Gamma$ where the pressure is applied, in this section we postpone making this
choice for generality.

On some parts of the boundary, $\Gamma_{u} \subset \Gamma$,
the displacement is fixed by a
Dirichlet boundary condition
\begin{eqnarray}
 u_i = \bar{u}_i,
\end{eqnarray}
while on other parts of the boundary $\Gamma_T \subset \Gamma$,
the traction is fixed by a Neumann boundary condition
\begin{eqnarray}
 n_i T_{ij} = T_{nj} = \bar{T}_{nj},
\end{eqnarray} where $n_i$ is normal to the boundary. 
Note that only one condition is applied at a given point on the boundary:
$\Gamma = \Gamma_u \cup \Gamma_T$ and $\Gamma_u \cap \Gamma_T = \emptyset$.

To find the weak form of Eq.~(\ref{eq:secondLaw}), begin by introducing a
virtual displacement vector $w_i$, with the property that $w_i=0$ on
$\Gamma_u$ (i.e., the virtual displacement vanishes on the Dirichlet
boundary). Otherwise, the $w_i$ are arbitrary functions of position. Contracting both sides of Eq.~(\ref{eq:secondLaw}) by $w_j$ and
integrating over the volume gives
\begin{eqnarray}
 0 & = & -\int_V dV w_j \nabla_i T_{ij} - \int_V dV w_j f_j.
\end{eqnarray} Integrating by parts gives
\begin{eqnarray}
 0 & = & \int_V dV \nabla_i\left(w_j\right) T_{ij}\nonumber\\
 & & - \int_V dV \nabla_i\left(w_j T_{ij}\right) - \int_V dV w_j f_j,
\end{eqnarray} which after applying Gauss's theorem becomes
\begin{eqnarray}
 0 & = & \int_V dV \nabla_i\left(w_j\right) T_{ij}\nonumber\\
 & & - \oint_\Gamma dA n_i w_j T_{ij} - \int_V dV w_j f_j.
\end{eqnarray} Because the virtual displacement $w_j$ vanishes on the Dirichlet
boundaries, the boundary term becomes
\begin{eqnarray}
\oint_\Gamma dA n_i w_j T_{ij} & = & \int_{\Gamma_{T}} dA n_i w_j T_{ij}, 
\end{eqnarray} so
\begin{eqnarray}
 0 & = & \int_V dV \nabla_i\left(w_j\right) T_{ij}\nonumber\\
 & & - \int_{\Gamma_{T}} dA w_j n_i T_{ij} - \int_V dV w_j f_j.
\end{eqnarray} Applying the Neumann boundary condition and inserting
Eqs.~(\ref{eq:Hooke}) and (\ref{eq:strain}) yields 
\begin{eqnarray}\label{eq:weak}
 0 & = & \int_V dV \nabla_i\left(w_j\right) Y_{ijkl} \nabla_k u_l
 \nonumber\\
 & & - \int_V dV w_j f_j - \int_{\Gamma_{T}} dA w_j \bar{T}_{nj}.
\end{eqnarray} This is the weak form of the elastostatic equations that we will use in the next subsection.

\subsection{Discretizing the weak form of the elastostatic equations}
We discretize Eq.~(\ref{eq:weak}) in a standard way. A similar derivation for the two-dimensional, amorphous, elastic equations is given in the discussion of deal.ii's step-8 tutorial~\cite{step8}, which solves the elasticity equations in two dimensions for amorphous materials.  For a detailed discussion of finite-element methods applied to the elasticity equations, see, e.g., Ref.~\cite{ibrahimbegovic2009nonlinear}.

To discretize Eq.~(\ref{eq:weak}), we choose $N$ scalar shape functions
$\phi_a\left(x_i\right)$, where $a=1,2,\ldots N$,
which are arbitrary functions of
position $x_i$. Then,
construct $3N$ three-dimensional vector shape functions, by
multiplying each scalar shape function by each of the $3$ orthonormal
basis vectors. E.g., one can define $\Phi_0 =
\left(\phi_0,0,0\right), \Phi_1 = \left(0,\phi_0,0\right), \Phi_2 =
\left(0,0,\phi_0\right), 
\Phi_3 = \left(\phi_1,0,0\right), \Phi_4 = \left(0,\phi_1,0\right), \Phi_5 = \left(0,0,\phi_1\right),\ldots$.
 
Expand the vectors $u_l$ and $w_j$ in terms of these vector-valued shape
functions 
\begin{eqnarray}
 u_l & = & U_A \Phi_{Al}\left(x_i\right),\\
 w_j & = & W_A \Phi_{Aj}\left(x_i\right).
\end{eqnarray} Here $\Phi_{Aj}$ is the $j^{th}$ vector component of the
$A^{th}$ vector shape function, and $U_A$ and $W_A$ are independent of
position $x_i$. Here and in the rest of this paper, the positional dependence
of the $\phi_A$ and $\Phi_{Ai}$ will be suppressed for clarity.

Inserting these expansions into
Eq.~(\ref{eq:weak}) gives
\begin{eqnarray}
0 & = & \int_V dV \nabla_i\left(W_A \Phi_{Aj} \right) Y_{ijkl} \nabla_k \left(U_B \Phi_{Bl}\right)
 \nonumber\\
 & & - \int_V dV W_A \Phi_{Aj} f_j - \int_{\Gamma_{T}} dA W_A\Phi_{Aj} \bar{T}_{nj}.
\end{eqnarray} Since $W_A$ and $U_A$ are independent of position, this becomes 
\begin{eqnarray}
0 & = & W_A U_B \int_V dV Y_{ijkl} \nabla_i \left(\Phi_{Aj}\right) \nabla_k \left(\Phi_{Bl}\right)
 \nonumber\\
 & & - W_A \int_V dV \Phi_{Aj} f_j - W_A \int_{\Gamma_{T}} dA \Phi_{Aj} \bar{T}_{nj}.
\end{eqnarray} We are free to choose shape functions that vanish on the
boundary, and since $w_j$ is arbitrary other than having to vanish on the boundary, we are free to choose $W_A=1$. This choice gives a discretized weak
form of the elastostatic equations suitable for solving via finite-element
methods:
\begin{eqnarray}
0 & = & U_B \int_V dV Y_{ijkl} \nabla_i \left(\Phi_{Aj}\right) \nabla_k \left(\Phi_{Bl}\right)
 \nonumber\\
 & & - \int_V dV \Phi_{Aj} f_j - \int_{\Gamma_{T}} dA \Phi_{Aj} \bar{T}_{nj}.
\end{eqnarray} Defining
\begin{eqnarray}
 M_{AB} \equiv \int_V dV Y_{ijkl} \nabla_i \left(\Phi_{Aj}\right) \nabla_k \left(\Phi_{Bl}\right),\\
 F_A \equiv \int_V dV \Phi_{Aj} f_j + \int_{\Gamma_{T}} dA \Phi_{Aj} \bar{T}_{nj},
\end{eqnarray} the equations can be written in matrix form as
\begin{eqnarray}
M_{AB} U_B = F_A.\label{eq:matrix}
\end{eqnarray}

In practice, the shape functions $\Phi_{Aj}$ are low-order
polynomials, with each function having support only within one
finite-element cell. We use second order polynomials for the shape functions except when we integrate to separately calculate the elastic energy in the substrate and coating, in which case we sub-divide each cell into 1000 smaller cells and interpolate with cubic-polynomial shape functions. 

We solve Eq.~(\ref{eq:matrix}) for
$U_B$ using deal.ii finite-element library~\cite{dealII82,dealII85}\footnote{We used version 8.2.1 of the dealII library to obtain the results in this paper, as it was current when we began developing our code. At the time of writing, deal.ii version 8.5 is available.}, the PETSc~\cite{petsc-web-page,petsc-user-ref,petsc_efficient} conjugate gradient linear solver, and the ParaSAILS
preconditioner in the Hypre package~\cite{falgout2002hypre}. Our specific implementation begins with the deal.ii step-8 tutorial~\cite{step8}, which solves the elastic equations in two dimensions for amorphous materials, and then generalizes to three dimensions and arbitrary Young's tensors (though in this paper, we only use isotropic and cubic-crystal Young's tensors).

\subsection{Mesh}
For simplicity, here we confine our attention to simple mirror geometries, generating our initial computational meshes using deal.ii's built-in primitives. We model the mirror as a simple cylinder.

We begin with an initial, coarse mesh. We create this coarse mesh by refining a deal.ii primitive mesh (cylinder or rectangular prism) in two stages: first, we apply two refinements to every element, and then we apply up to two additional refinements on elements within one beam width $r_0$ of the mirror's front, center point. Then, after achieving a solution on the coarse mesh, we refine using adaptive mesh refinement. We estimate the numerical error in each cell using the Kelly error estimator~\cite{kelly1983posteriori,gago1983posteriori}, and then we rank them by this error estimate. We then refine the top 14\% and coarsen the bottom 2\%, approximately doubling the number of cells with each refinement\footnote{When running on multiple processors (i.e., multiple cores), we divide the mesh among them. We then refine the top 14\% and coarsen the bottom 2\% of cells on each processor.}.

\begin{figure*}[ht]
\includegraphics[width=6.5in]{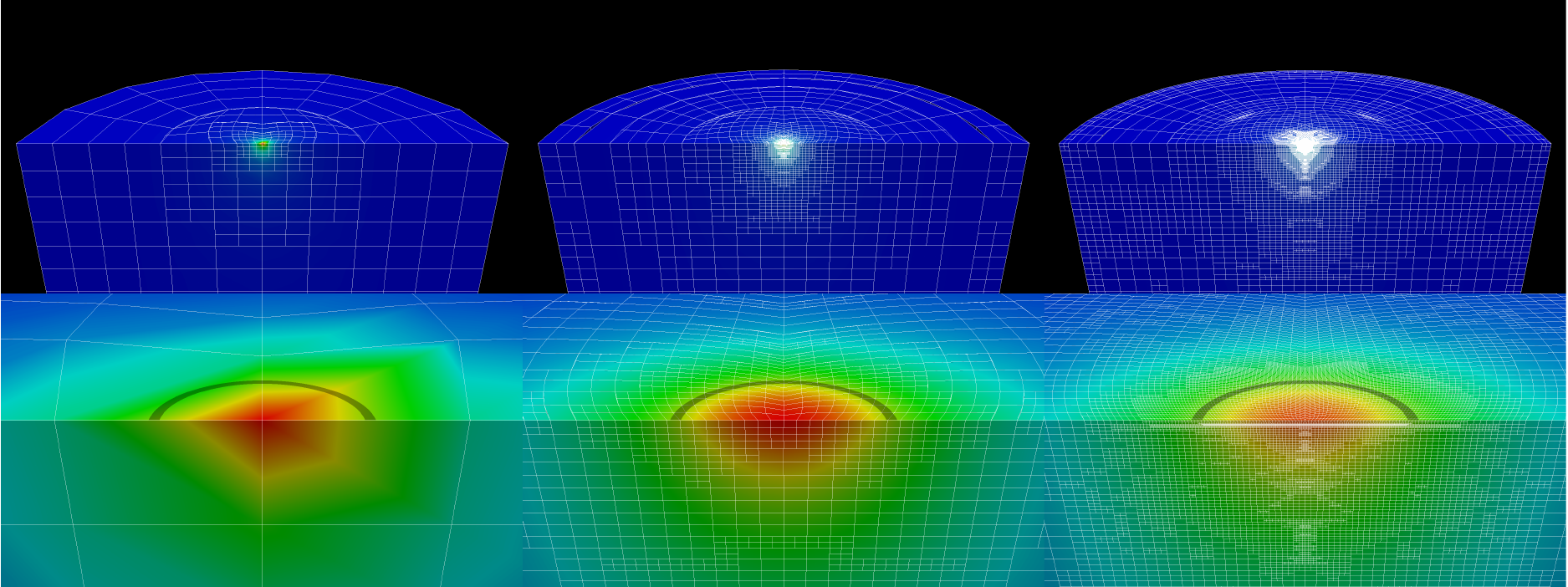}
\caption{
A cross-sectional view of a sample meshed cylindrical domain (fused silica mirror with a cubic crystalline coating) that has been refined, from left to right, 1, 5, and 9 times using adaptive mesh refinement. Below each upper panel, a lower panel zooms in. The dark ring in each image has radius $r_0$ to represent the Gaussian profile pressure applied to the cylinder. The magnitude of the resulting displacement is shown by the coloring. \label{fig:mesh}}
\end{figure*}

\subsection{Performance}\label{sec:performance}

\begin{figure*}[ht]
\begin{center}
\includegraphics[width=4.0in]{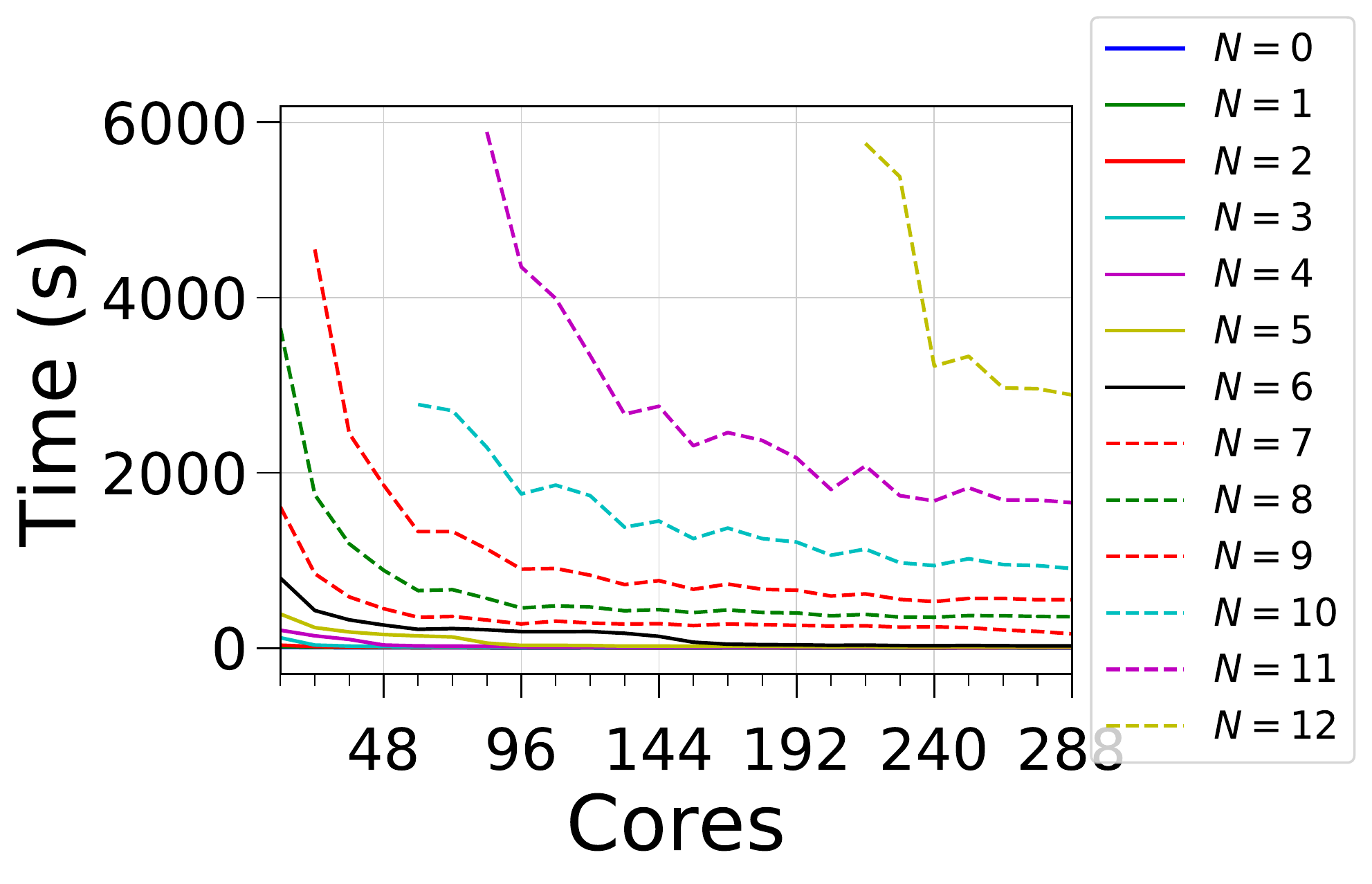}
\caption{
The run time of our thermal-noise calculations as a function of the number of compute cores used in the calculation. Colors indicate resolution $N$, where each resolution $N$ has approximately twice the number of finite elements as resolution $N-1$. Performance improves with increasing numbers of cores until it levels off; we suspect this is caused by increasing communication costs. Running on more cores (i.e., on more compute nodes) increases the available memory. Given the memory limits of the cluster where we ran these calculations, the highest resolutions achieved ($N>=9$) require running on more cores (and thus on more compute nodes and more total memory). \label{fig:allcycles}} 
\end{center}
\end{figure*}

We have tested how the performance of our code varies with increasing resolution and increasing numbers of compute cores. We label each resolution by an integer $N$, where increasing $N$ by one corresponds to approximately doubling the number of finite-element cells. Figure~\ref{fig:allcycles} shows the wall-clock time elapsed for each resolution of a typical thermal-noise calculation. Initially, increasing the number of processors decreases the time required to complete a given resolution; however, the performance then hits a plateau, as communication costs increase.

For the highest resolutions (e.g., $N=12$ in Fig.~\ref{fig:allcycles}), we often run on more cores than Fig.~\ref{fig:allcycles} would suggest are necessary, because these high resolutions (with hundreds of millions of finite-element degrees of freedom) require the memory from the additional compute nodes. Also, note that occasionally, when performing these timing tests, we observed spuriously inconsistent timing for some simulations; we suspect this occurred when our cluster's network became saturated.



\subsection{Analytic approximate solutions}
The amplitude spectral density of the substrate thermal noise $\sqrt{S_q}$, assuming a semi-infinite amorphous mirror with Young's modulus $Y_{\rm sub}$, mechanical loss angle $\phi_{\rm sub}$, and Gaussian beam radius $r_0$, is given by the square root of its power spectral density (e.g., Eq.~(59) of. Ref.~\cite{liu2000thermoelastic}):
\begin{eqnarray}
\sqrt{S_q} & = & \sqrt{\frac{\sqrt{2}k_BT}{\pi^{3/2}f}\frac{1-\sigma_{\rm sub}^2}{Y_{\rm sub} r_0}\phi_{\rm sub}}.\label{eq:analyticSubstrate}
\end{eqnarray}
For an amorphous substrate with a thin, reflective, amorphous coating, the coating thermal noise is given by (e.g., Eq.~(21) of Ref.~\cite{harry2002thermal} with $\phi_{\|}=\phi_\bot=\phi_{\rm coat}$)
\begin{eqnarray}
\sqrt{S_q} & = & \sqrt{\frac{k_BT}{\pi^2f}\frac{1-\sigma_{sub}^2}{r_0 Y_{\rm sub}}\frac{d}{r_0}\frac{\phi_{\rm coat}}{Y_{\rm sub}Y_{\rm coat}(1-\sigma_{\rm coat}^2)(1-\sigma_{\rm sub}^2)}}\nonumber\\
& \times & \sqrt{Y_{\rm coat}^2(1+\sigma_{\rm sub})^2(1-2\sigma_{\rm sub})^2+Y_{\rm sub}^2(1+\sigma_{\rm coat})^2(1-2\sigma_{\rm coat})}\label{eq:analyticCoat}
\end{eqnarray}
Here, to facilitate comparison with our numerical calculations, we treat the coating as having a single mechanical loss angle $\phi$. More realistically, an amorphous coating should have one loss angle for each of its two independent elastic moduli~\cite{hong2013brownian} of this paper), and a cubic crystalline coating should have one loss angle for each of the three independent components in its Young's tensor $Y_{ijkl}$ (cf. the discussion in Sec.~(\ref{sec:lossangles}).

Note that because we are considering noise in a single mirror, rather than two (as in Ref.~\cite{cole2013tenfold}), our analytic formula differs from Eq.~(1) of Ref.~\cite{cole2013tenfold}.

\section{Results}\label{sec:results}
Unless stated otherwise (e.g., when adjusting a parameter to observe how the numerical solution scales with the adjustments), the numerical parameters in our calculations are taken from Table~\ref{tab:params}. To take a concrete example, we choose the parameters in the top two sections of Table~\ref{tab:params} to agree with those in the supplementary information for Ref.~\cite{cole2013tenfold}.  

When we treat the reflective coating as a crystal, we use the elastic moduli $c_{11}$, $c_{12}$, and $c_{44}$ given in the middle section of Table 1. 
Also following Ref.~\cite{cole2013tenfold}, we consider the specific case of ${\rm Al}_x{\rm Ga}_{1-x}As$ with $x=0.92$, and we take the values of the elastic moduli from Sec.~S2 of the supplementary information of Ref.~\cite{gehrsitz1999compositional}. We use the same loss angle as in the effective isotropic case. Note that in our numerical calculations, we choose a cubic crystal lattice orientation so that the cube faces are parallel to the mirror faces. We also use the same, single loss angle as in the effective isotropic case. 

The bottom section of Table~\ref{tab:params} shows our choice for the mirror radius, where we choose a ``typical'' height that gives a mirror diameter of approximately 1 inch. We choose the total height of the mirror (including the coating) to be the sum of the mirror radius and the coating thickness. 

The amplitude of the thermal noise, $\sqrt{S_q}(f)$, is proportional to $f^{-1/2}$ [Eq.~(\ref{eq:FDT})]. For concreteness, we evaluate the thermal noise at a frequency $f=100\mbox{ Hz}$, chosen as a representative frequency of where Advanced LIGO is most sensitive (cf. Fig.~\ref{fig:gwinc}).

\begin{table}
\begin{center}
\begin{tabular}{|c|c|c|}
\hline
Parameter & Description & Value\\
\hline
$r_0$ & beam width & 177 $\mathrm{\mu m}$\\
$d$ & coating thickness & 6.83 $\mathrm{\mu m}$\\
$T$ & temperature & 300 K\\
$\sigma_{\rm SiO_2}$ & fused silica Poisson ratio &  0.17\\
$\sigma_{\rm iso}$ & effective isotropic AlGaAs Poisson ratio &  0.32\\
$Y_{\rm SiO_2}$ & fused silica Young's modulus &  72 GPa\\
$Y_{\rm iso}$ & effective isotropic AlGaAs Young's modulus &  100 GPa\\
$\phi_{\rm SiO_2}$ & fused silica loss angle &  1$\times 10^{-6}$\\
$\phi_{\rm iso}$ & effective isotropic AlGaAs loss angle &  2.5$\times 10^{-5}$\\
\hline
$c_{11}$ & ${\rm Al}_{x}{\rm Ga}_{1-x}{\rm As}$ elastic modulus & 119.94 GPa\\
$c_{12}$ & ${\rm Al}_{x}{\rm Ga}_{1-x}{\rm As}$ elastic modulus & 55.38 GPa\\
$c_{44}$ & ${\rm Al}_{x}{\rm Ga}_{1-x}{\rm As}$ elastic modulus & 59.15 GPa\\
$x$ & $Al_{x}Ga_{1-x}$ As fraction of aluminum  & 0.92\\
$\phi_{\rm crystal}$ & crystalline AlGaAs loss angle &  2.5$\times 10^{-5}$\\
\hline
$R$ & mirror radius & 12500 $\mathrm{\mu m}$\\
$H$ & mirror height & $d+R$ $\mathrm{\mu m}$\\
$f$ & frequency &  100 Hz\\
\hline
\end{tabular}
\caption{Numerical values of parameters used in our numerical thermal noise computations, unless otherwise stated (e.g., when adjusting a parameter to check the expected scaling). \label{tab:params}}
\end{center}
\end{table}

\begin{figure*}[ht]
\includegraphics[width=3.15in]{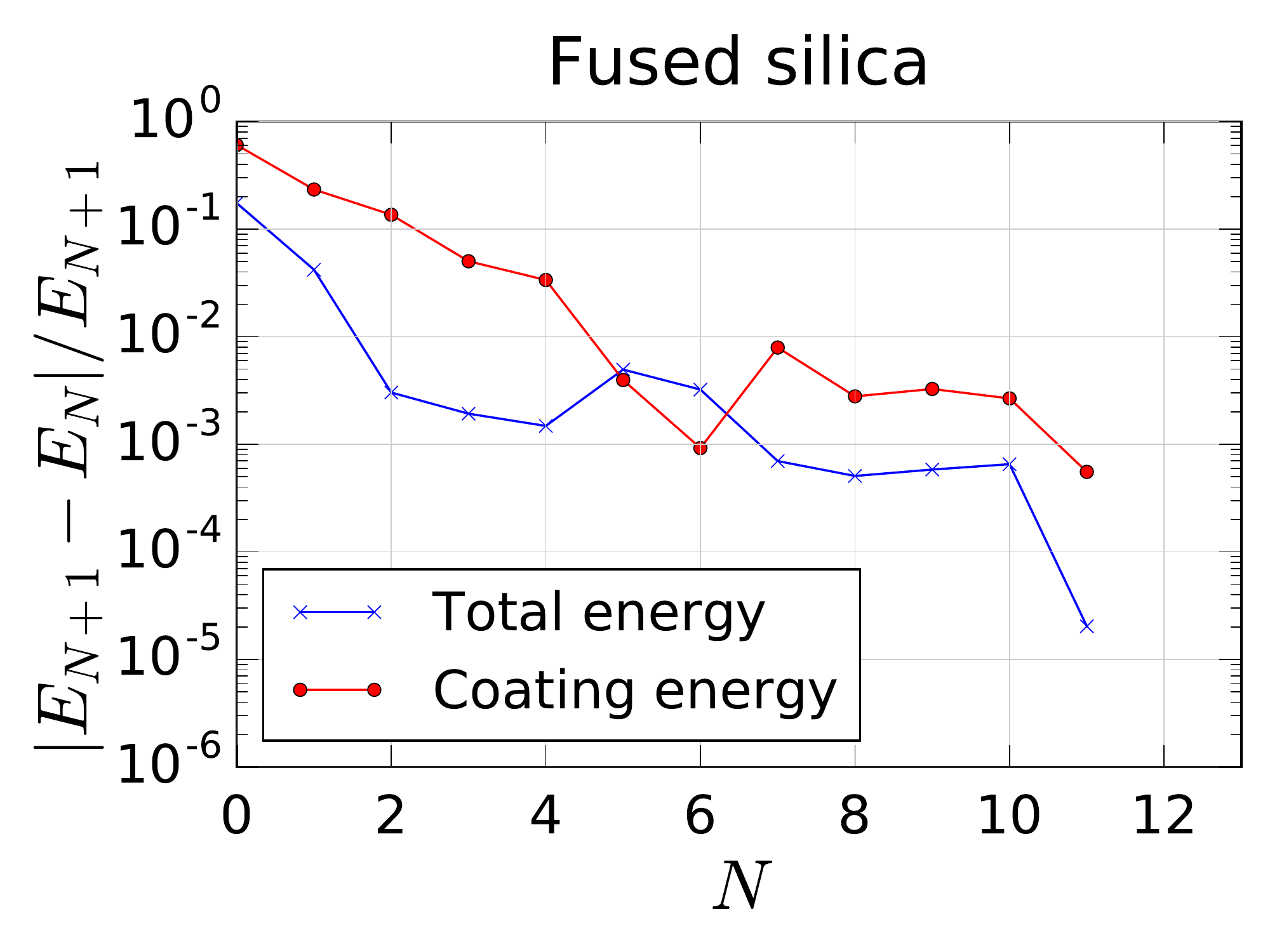}
\\
\includegraphics[width=3.15in]{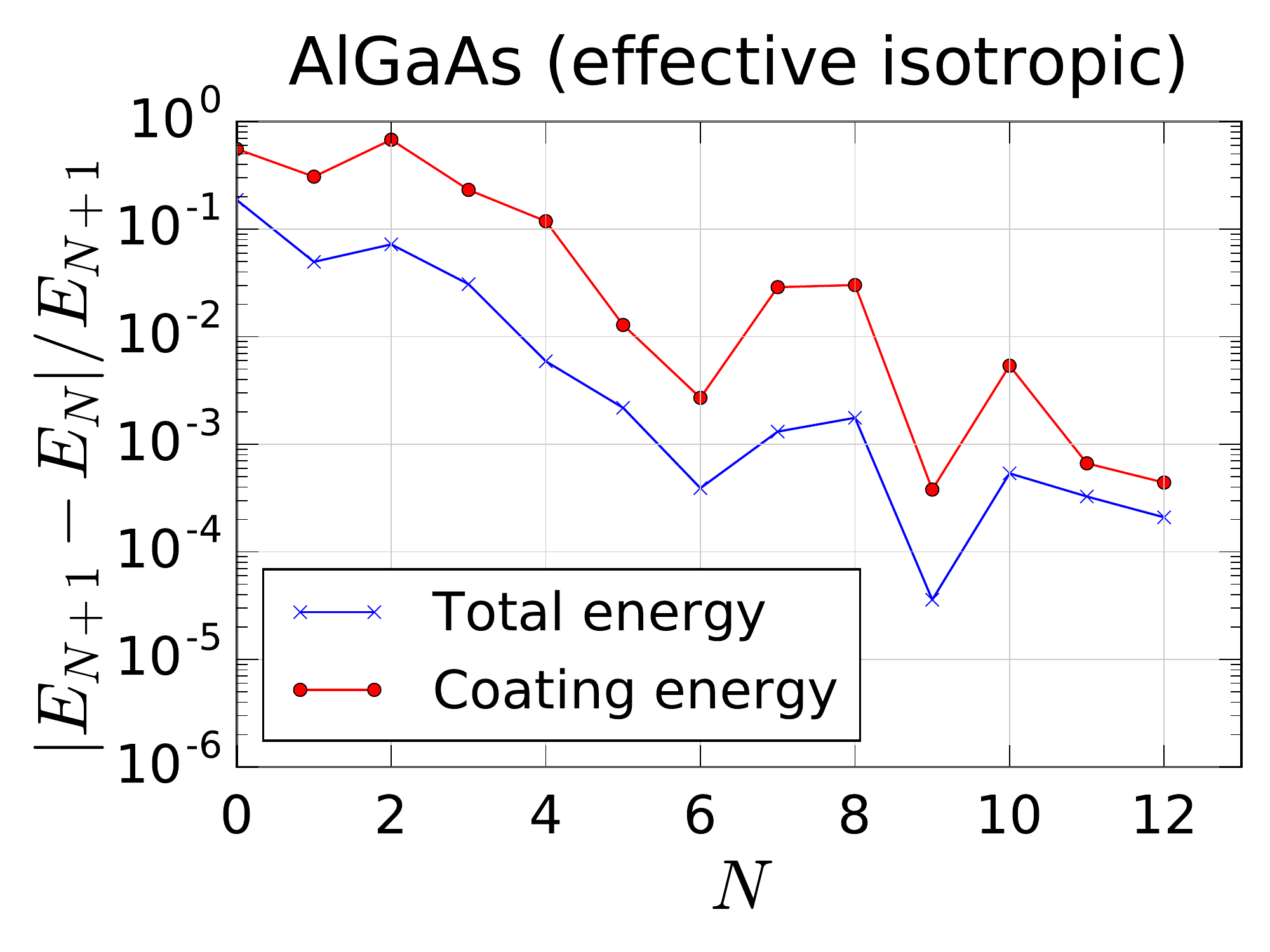}
\includegraphics[width=3.15in]{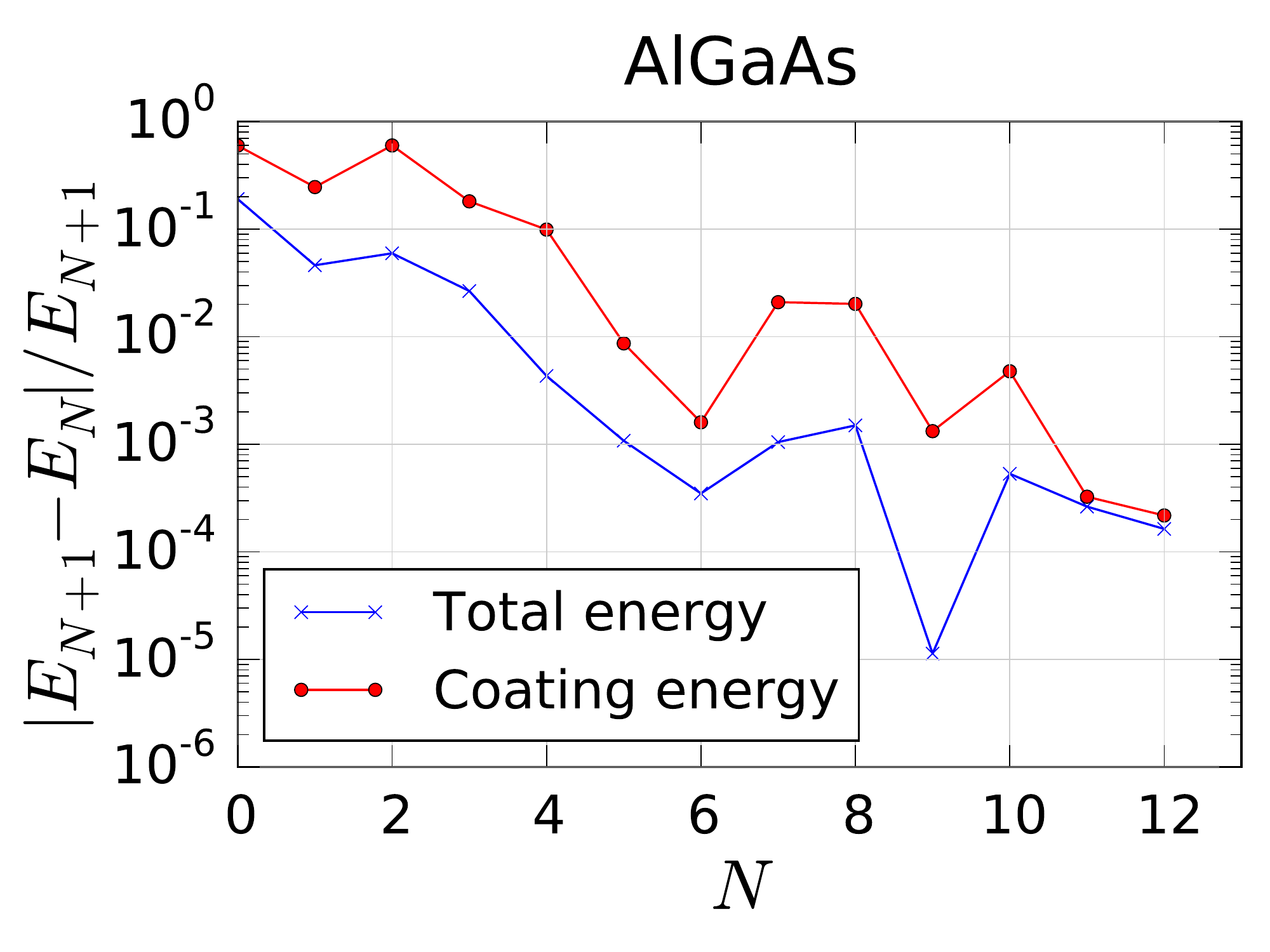}
\caption{
Numerical convergence for a fused silica mirror with various coatings: \emph{Top:} fused silica, \emph{Bottom Left:} AlGaAs (effective isotropic), \emph{Bottom Right:} AlGaAs (crystalline) \label{fig:convergence}}
\end{figure*}

Figure~\ref{fig:convergence} assesses our code's numerical convergence by showing fractional differences in the total stored energy and in the coating stored energy as a function of resolution $N$ (Cf. Sec.~\ref{fig:mesh}). In the first plot, we consider a mirror made entirely of fused silica, treating a thin slice of thickness $d$ on the top face of the cylinder as the coating. The remaining panels in Fig.~\ref{fig:convergence} use an AlGaAs coating, with the first treating the coating as an effective isotropic material (with effective bulk and shear elastic moduli) and with the second treating the coating as a cubic crystal with three independent elastic modulus components. We do not expect perfectly monotonic behavior, because adaptive mesh refinement does not uniformly increase resolution. Nevertheless, we are satisfied that the differences are generally decreasing until they achieve a fractional error of 0.1\%, which is sufficient for our purposes (e.g., sufficient to compare with experimental results).

\begin{figure*}[ht]
\includegraphics[width=3.15in]{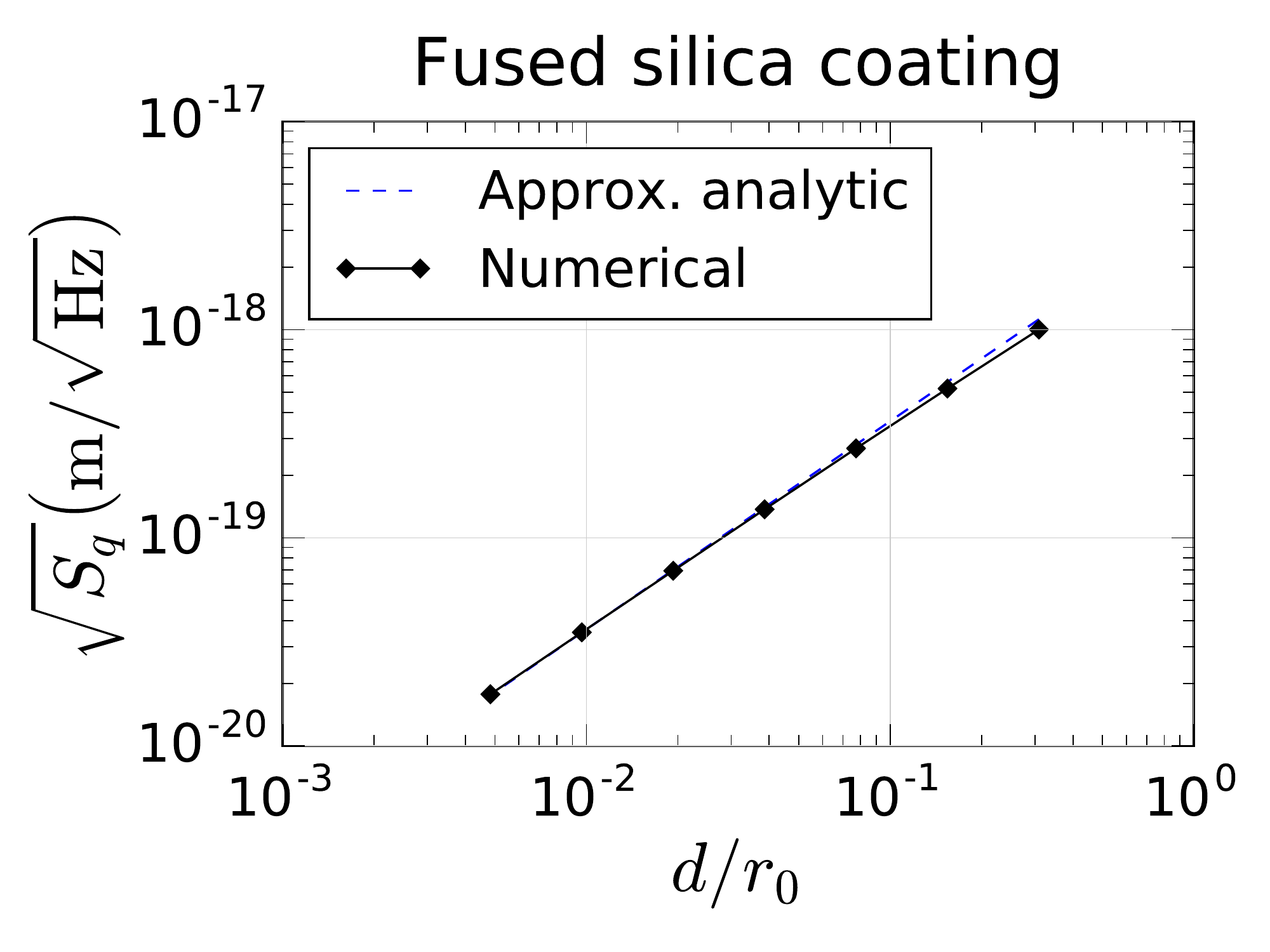}
\includegraphics[width=3.15in]{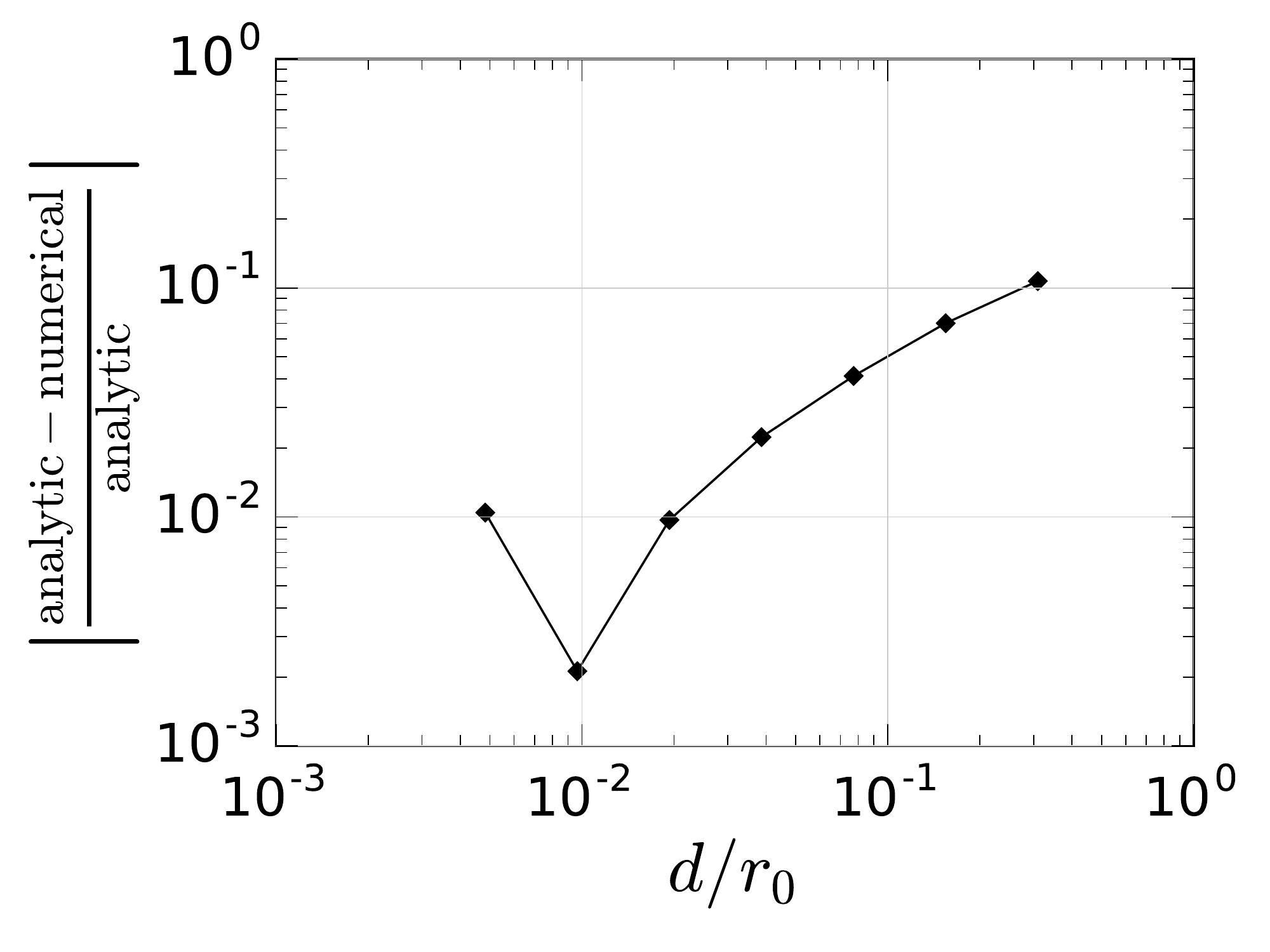}
\includegraphics[width=3.15in]{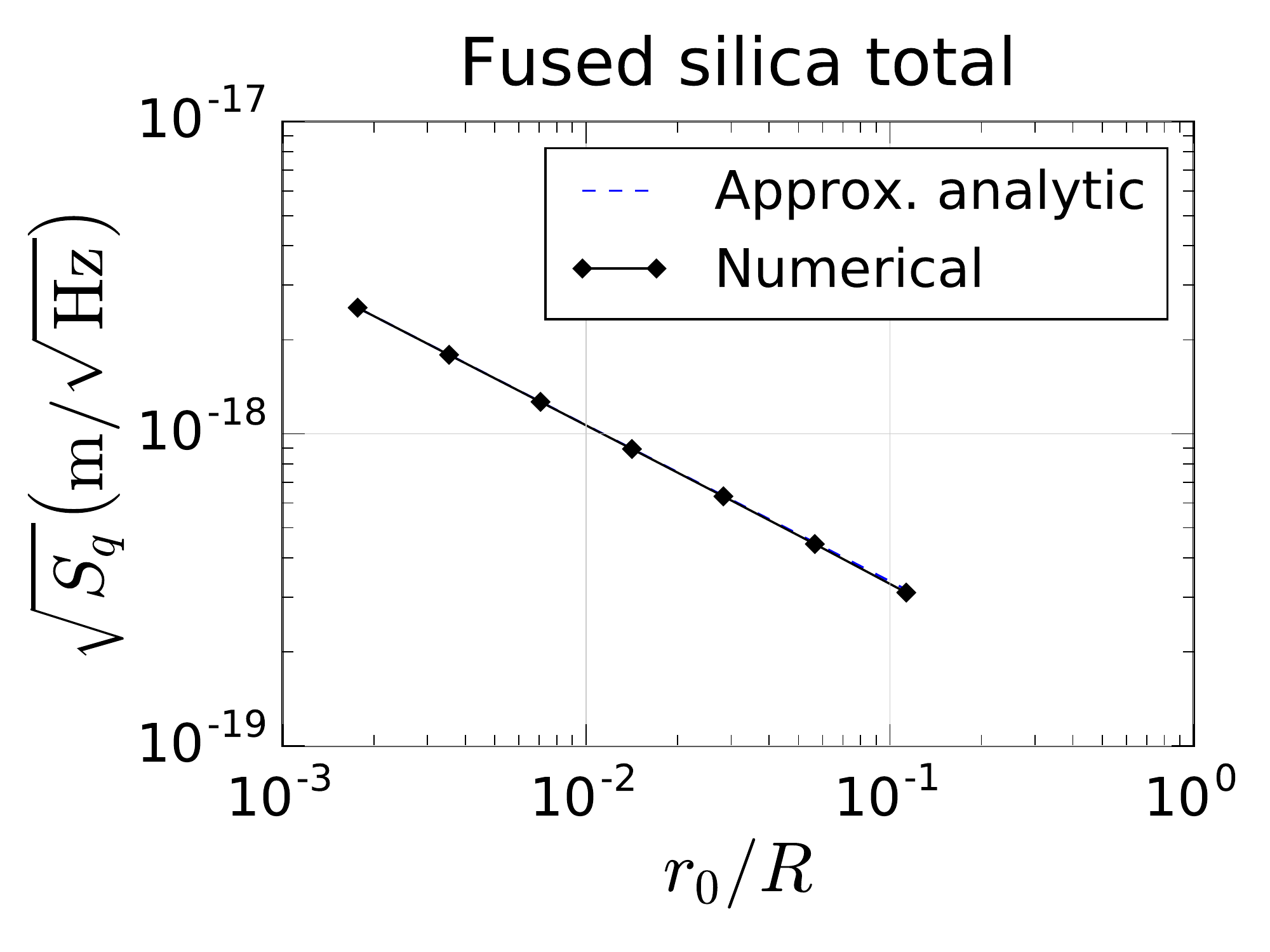}
\includegraphics[width=3.15in]{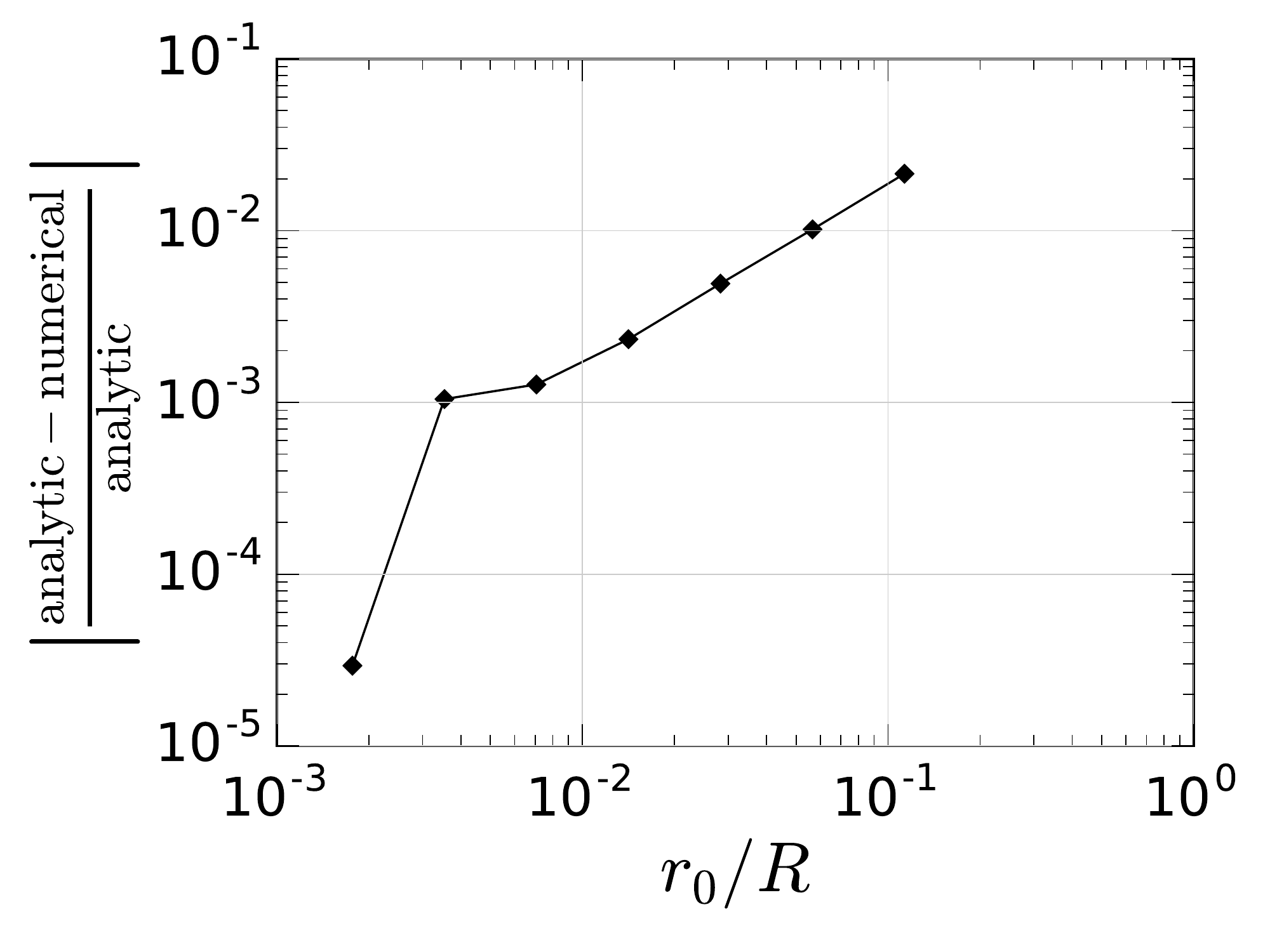}
\includegraphics[width=3.15in]{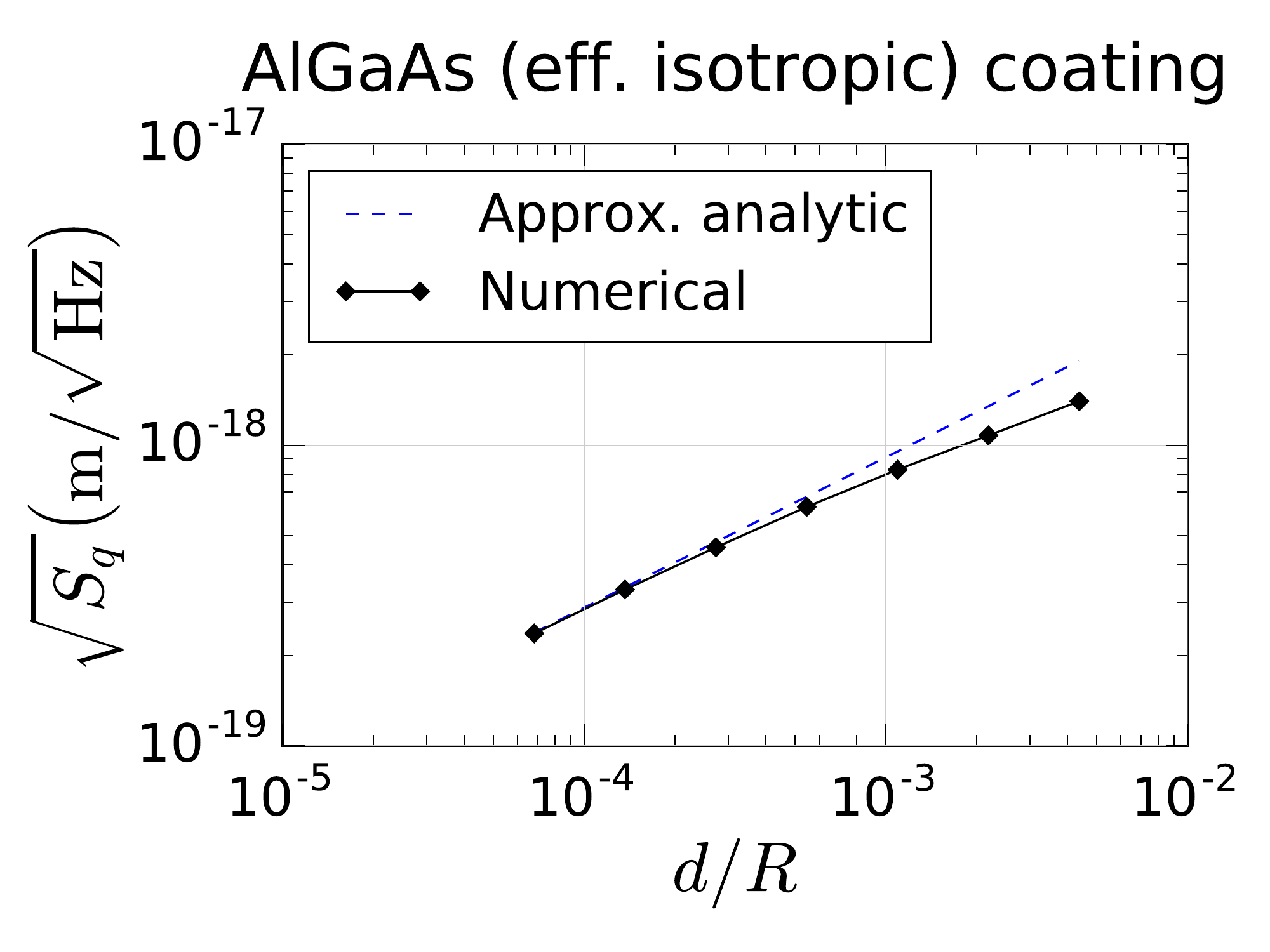}
\includegraphics[width=3.15in]{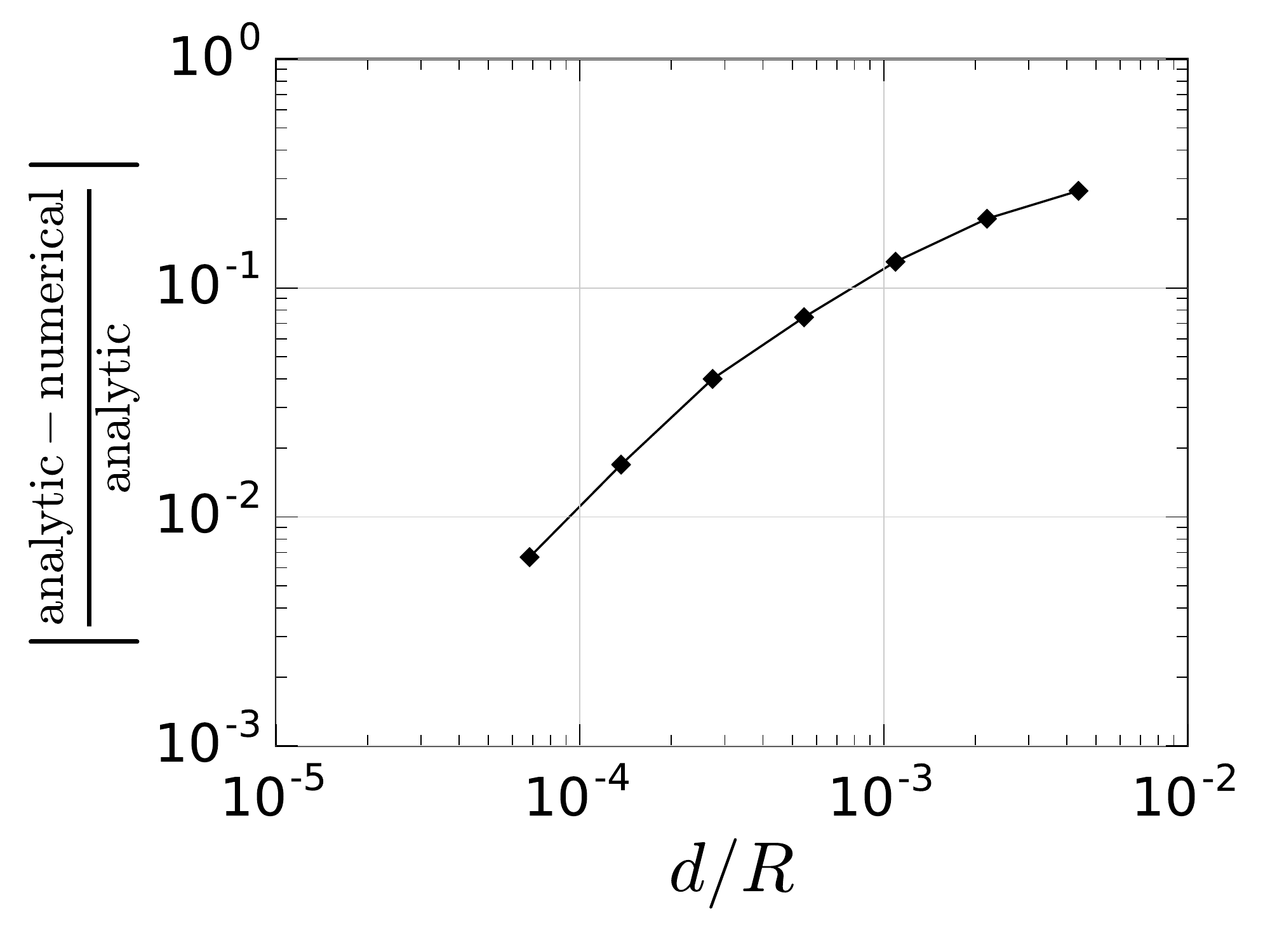}
\caption{
Numerical computations of thermal noise for different mirrors. The left column shows the numerical thermal noise and the approximate analytic solutions given in Eqs.~(\ref{eq:analyticSubstrate})--(\ref{eq:analyticCoat}). The right column shows the fractional differences between the numerical and approximate analytic solutions. The top two rows show thermal noise for a mirror where both the substrate and coating are made of fused silica, while the bottom row shows thermal noise for a fused-silica substrate with a AlGaAs coating, treated as an effective isotropic material (so that the analytic solutions can be used).\label{fig:edge}}
\end{figure*}

Figure~\ref{fig:edge} compares our code's results to known, approximate, analytic solutions, for isotropic, amorphous coatings. We find that our numerical solutions approach the expected analytic solutions in the appropriate limits. In Fig.~\ref{fig:edge}, the left panels show thermal noise as a function of different dimensionless ratios that each characterize a different approximation in the analytic solutions. The right panels show differences between the numerical and approximate analytic results. Each numerical point represents the highest simulated resolution for the given physical dimensions. 

In the top two rows of Fig.~\ref{fig:edge}, we show the coating thermal noise for different beam sizes $r_0$, holding all other quantities fixed. The mirror is entirely fused silica, with the topmost layer of thickness $d$ treated as the coating. In the top row, we show the coating thermal noise as a function of the dimensionless quantity $d/r_0$, which is small in the thin-coating approximation used in the analytic solution. The numerical solution approaches the analytic as $d/r_0$ approaches zero, as expected. In the middle row, we show the total thermal noise for different beam widths $r_0$ as a function of the dimensionless quantity $r_0/R$, which characterizes the importance of edge effects, which are neglected in the analytic solution (i.e., the analytic solution does not depend on $R$). The mirror is again entirely fused silica. Again, the numerical solution approaches the analytic as $r_0/R$ approaches zero, as expected. 

In the bottom row of Fig.~\ref{fig:edge}, we show the coating thermal noise for different coating thicknesses $d$ as a function of the dimensionless quantity $d/R$, which characterizes the thin-coating approximation. The mirror in this case is made of two materials, a fused silica substrate of thickness $R$ and an effective isotropic AlGaAs coating of thickness $d$. The numerical solution approaches the analytic as $d/R$ approaches zero, as expected. Finally, note that the value of each point in the difference plots are within the numerical error of that particular point's simulation. Larger numerical error (caused, e.g., by greater difficulty in resolving different length scales) explains the anomalous behavior of the left most points.

\begin{figure*}[ht]
\begin{center}
\includegraphics[width=3.15in]{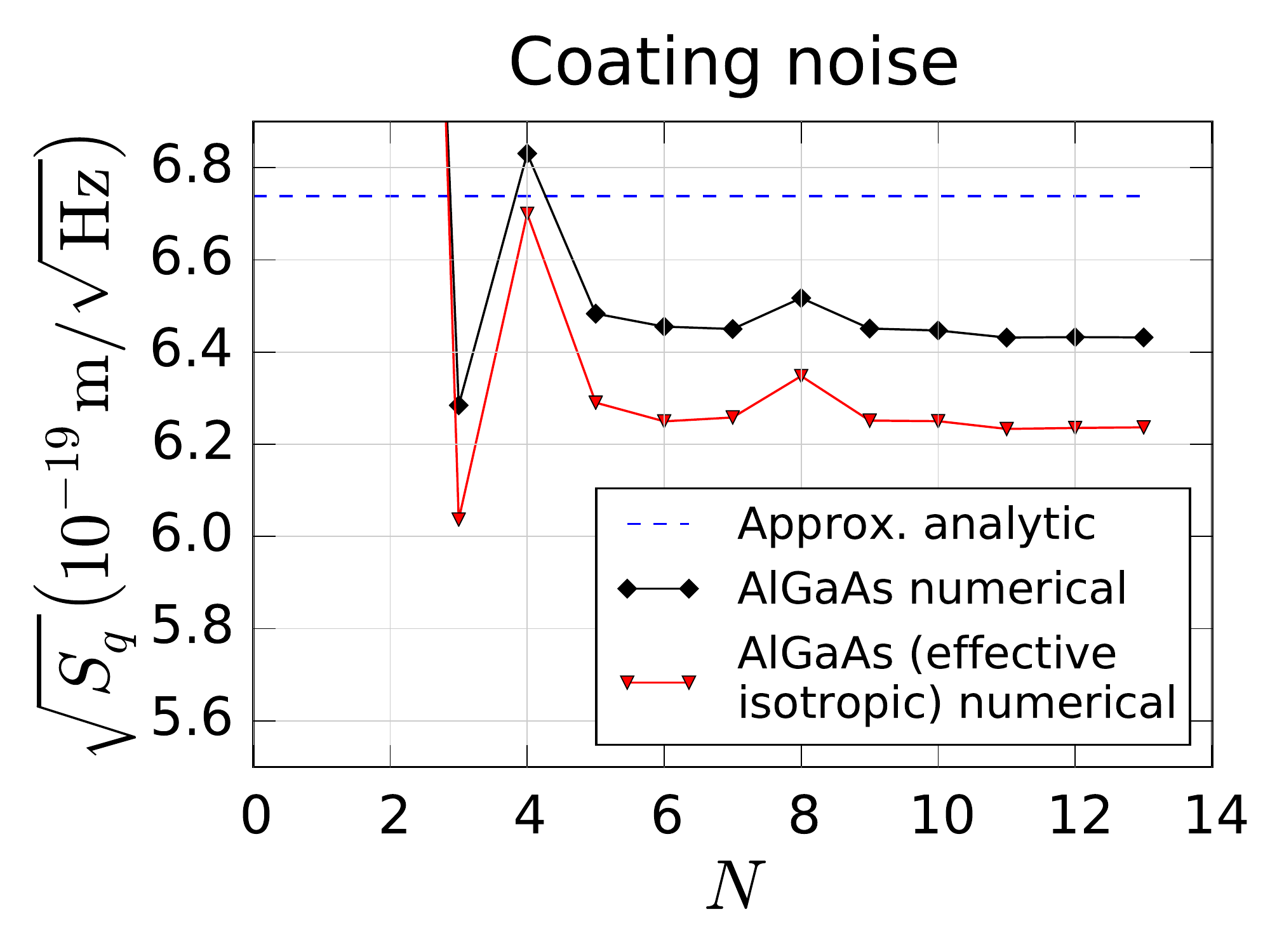}
\caption{
Coating Brownian noise at 
$f=100\mbox{ Hz}$ for analytic and numerical effective isotropic and numerical anisotropic crystalline elastic moduli. \label{fig:noise}}
\end{center}
\end{figure*}

Finally, we compare the numerical thermal noise for AlGaAs coatings on fused silica substrates, comparing the results when the coating is treated as a cubic crystal in the elastic problem to results treating the coating as an effective isotropic material. Figure~\ref{fig:noise} shows the coating thermal noise as a function of resolution for two mirrors with the fused silica substrate: one with an AlGaAs effective isotropic coating and one with an AlGaAs crystalline coating. We compare both numerical results to an approximate analytic solution for an amorphous, semi-infinite mirror with a thin, amorphous coating. We resolve the effect of treating the crystalline coating as a cubic crystal. 

Unlike the analytic solution, neither numerical solution neglects edge effects or coating thickness effects in the elasticity calculation. As a result of these finite-size effects, the effective isotropic and analytic solutions differ by approximately 7\%. Additionally including crystalline effects (i.e., treating the coating as a cubic crystal, rather than as an amorphous material) causes the thermal noise to differ from the approximate, analytic solution by about 4\%, since the crystalline numerical result is about 3\% larger than the effective-isotropic numerical result. For the particular case we consider (mirror dimension, beam size, temperature, etc.), then, we conclude that including finite-size effects causes a deviation from the approximate, analytic solution comparable in magnitude to that caused by treating the coating as a cubic crystal. 

\section{Conclusion}
\label{sec:conclusion}

In this paper, we have numerically computed the Brownian substrate and coating thermal noise for a cylindrical mirror with a thin, reflective, possibly crystalline coating. To do this, we have developed a new tool, built on open-source libraries, that computes thermal noise by solving an elastostatic problem and inserting the solution into the fluctuation-dissipation theorem. Using a parallel finite-element method with adaptive mesh refinement, we have demonstrated numerical convergence, resolutions up to approximately $2 \times 10^{8}$ degrees of freedom, and the capability to run on up to hundreds of processors. Because of limited memory on the cluster where we performed our calculations, the highest resolutions were only achievable with the increased memory available with running on a larger number of processors than would otherwise be necessary. 

Using this new tool, we have computed the Brownian thermal noise for a cylindrical mirror with a thin, reflective coating. When the coating is amorphous, we agree well with approximate, analytic solutions that neglect edge effects and anisotropic effects, and our numerical results scale as expected with beam radius, mirror size, and coating thickness. When the coating is a cubic crystal (specifically, AlGaAs), our numerical results show a small but significant difference between the noise computed accounting for the crystal's anisotropy and the noise computed while treating the crystal as an effective isotropic material. The C++ source code for our tool is available at \cite{thermalNoiseRepo}. 

Because our code is open, additional physics can be incorporated in future work, supporting the long-term goal of understanding and reducing thermal noise in future gravitational-wave detectors. For instance, as discussed in Sec.~\ref{sec:lossangles}, rather than using a single mechanical loss angle, one could extend our tool to treat the mechanical loss in the coating more realistically, by introducing one loss angle for each independent component of the Young's tensor, generalizing the ``bulk'' and ``shear'' loss angles introduced in Ref.~\cite{hong2013brownian}. Other directions for future work include extending our code to compute thermoelastic and thermorefractive noise and incorporating more realistic LIGO mirror shapes with physically correct edge effects. The mirrors could also be treated more realistically, e.g., by including a more realistic mirror shape (including ``ears'' that are held fixed by suspension fibers~\cite{heptonstall2014}) and adjusting our outer boundary condition accordingly. Another potential direction for future study includes varying the laser beam intensity profile. Flatter intensity profiles better average thermal fluctuations than Gaussian profiles do; one can use our tool to explore numerically how thermal noise varies with beam shape, generalizing results for semi-infinite, amorphous optics~\cite{o2004implications,
braginsky2003thermodynamical,fejer2004thermoelastic,vinet2005mirror,
o2006note,Lovelace:2006ne} to crystalline optics of finite size. Finally, realistic LIGO optics use multi-layer coatings; improved, high-accuracy meshes could potentially enable our tool to explore the effects of multiple layers on the coating thermal noise. 
\\
\ack{We are pleased to acknowledge Rana X.~Adhikari, Garrett Cole, and Joshua R.~Smith for helpful discussions. This work was supported in part by National
 Science Foundation grants PHY-1307489, PHY-1606522, PHY-1654359, and AST-1559694. Computations in this paper were performed on the Orange County Relativity cluster for Astronomy (ORCA), which is supported in part by National Science Foundation grant PHY-1429873, by the Research Corporation for Science
 Advancement, and by California State University, Fullerton. Some of the 
formulas in this paper were checked using Mathematica.}

\section*{References}
\bibliographystyle{iopart_num}
\bibliography{References2,References3,References4}

\providecommand{\newblock}{}
\begin{thebibliography}{10}
\expandafter\ifx\csname url\endcsname\relax
  \def\url#1{{\tt #1}}\fi
\expandafter\ifx\csname urlprefix\endcsname\relax\def\urlprefix{URL }\fi
\providecommand{\eprint}[2][]{\url{#2}}

\bibitem{Abbott:2016blz}
{{B~P~Abbott et al~for the LIGO and Virgo Scientific Collaborations}} (Virgo,
  LIGO Scientific) 2016 {\em Phys. Rev. Lett.\/} {\bf 116} 061102
  (\textit{Preprint} \eprint{1602.03837})

\bibitem{Abbott:2016nmj}
{B~P~Abbott et al~for the LIGO and Virgo Scientific Collaborations} 2016 {\em
  Phys. Rev. Lett.\/} {\bf 116} 241103 (\textit{Preprint} \eprint{1606.04855})

\bibitem{TheLIGOScientific:2016pea}
{B~P~Abbott et al~for the LIGO Scientific Collaboration} 2016 {\em Physical
  Review X\/} {\bf 6} 041015

\bibitem{scientific2017gw170104}
{B~P~Abbott et al for the LIGO Scientific Collaboration and the Virgo
  Collaboration} 2017 {\em Physical Review Letters\/} {\bf 118} 221101

\bibitem{TheVirgo:2014hva}
Acernese F, Agathos M, Agatsuma K, Aisa D, Allemandou N, Allocca A, Amarni J,
  Astone P, Balestri G, Ballardin G, Barone F, Baronick J~P, Barsuglia M, Basti
  A, Basti F, Bauer T~S, Bavigadda V, Bejger M, Beker M~G, Belczynski C,
  Bersanetti D, Bertolini A, Bitossi M, Bizouard M~A, Bloemen S, Blom M, Boer
  M, Bogaert G, Bondi D, Bondu F, Bonelli L, Bonnand R, Boschi V, Bosi L,
  Bouedo T, Bradaschia C, Branchesi M, Briant T, Brillet A, Brisson V, Bulik T,
  Bulten H~J, Buskulic D, Buy C, Cagnoli G, Calloni E, Campeggi C, Canuel B,
  Carbognani F, Cavalier F, Cavalieri R, Cella G, Cesarini E, Chassande-Mottin
  E, Chincarini A, Chiummo A, Chua S, Cleva F, Coccia E, Cohadon P~F, Colla A,
  Colombini M, Conte A, Coulon J~P, Cuoco E, Dalmaz A, D'Antonio S, Dattilo V,
  Davier M, Day R, Debreczeni G, Degallaix J, Del\'{e}glise S, Pozzo W~D,
  Dereli H, Rosa R~D, Fiore L~D, Lieto A~D, Virgilio A~D, Doets M, Dolique V,
  Drago M, Ducrot M, Endr\"{o}czi G, Fafone V, Farinon S, Ferrante I, Ferrini
  F, Fidecaro F, Fiori I, Flaminio R, Fournier J~D, Franco S, Frasca S,
  Frasconi F, Gammaitoni L, Garufi F, Gaspard M, Gatto A, Gemme G, Gendre B,
  Genin E, Gennai A, Ghosh S, Giacobone L, Giazotto A, Gouaty R, Granata M,
  Greco G, Groot P, Guidi G~M, Harms J, Heidmann A, Heitmann H, Hello P,
  Hemming G, Hennes E, Hofman D, Jaranowski P, Jonker R~J~G, Kasprzack M,
  K\'{e}f\'{e}lian F, Kowalska I, Kraan M, Kr\'{o}lak A, Kutynia A, Lazzaro C,
  Leonardi M, Leroy N, Letendre N, Li T~G~F, Lieunard B, Lorenzini M, Loriette
  V, Losurdo G, Magazz\`{u} C, Majorana E, Maksimovic I, Malvezzi V, Man N,
  Mangano V, Mantovani M, Marchesoni F, Marion F, Marque J, Martelli F,
  Martellini L, Masserot A, Meacher D, Meidam J, Mezzani F, Michel C, Milano L,
  Minenkov Y, Moggi A, Mohan M, Montani M, Morgado N, Mours B, Mul F, Nagy M~F,
  Nardecchia I, Naticchioni L, Nelemans G, Neri I, Neri M, Nocera F, Pacaud E,
  Palomba C, Paoletti F, Paoli A, Pasqualetti A, Passaquieti R, Passuello D,
  Perciballi M, Petit S, Pichot M, Piergiovanni F, Pillant G, Piluso A, Pinard
  L, Poggiani R, Prijatelj M, Prodi G~A, Punturo M, Puppo P, Rabeling D~S,
  R\'{a}cz I, Rapagnani P, Razzano M, Re V, Regimbau T, Ricci F, Robinet F,
  Rocchi A, Rolland L, Romano R, Rosi\'{n}ska D, Ruggi P, Saracco E, Sassolas
  B, Schimmel F, Sentenac D, Sequino V, Shah S, Siellez K, Straniero N,
  Swinkels B, Tacca M, Tonelli M, Travasso F, Turconi M, Vajente G, van Bakel
  N, van Beuzekom M, van~den Brand J~F~J, Broeck C~V~D, van~der Sluys M~V, van
  Heijningen J, Vas\'{u}th M, Vedovato G, Veitch J, Verkindt D, Vetrano F,
  Vicer\'{e} A, Vinet J~Y, Visser G, Vocca H, Ward R, Was M, Wei L~W, Yvert M,
  \.{z}ny A~Z and Zendri J~P (VIRGO) 2015 {\em Class.\ Quantum Grav.\/} {\bf
  32} 024001 (\textit{Preprint} \eprint{1408.3978})

\bibitem{Somiya:2011np}
Somiya K (KAGRA Collaboration) 2012 {\em Class.\ Quantum Grav.\/} {\bf 29}
  124007 (\textit{Preprint} \eprint{1111.7185})

\bibitem{aso2013interferometer}
Aso Y, Michimura Y, Somiya K, Ando M, Miyakawa O, Sekiguchi T, Tatsumi D and
  Yamamoto H 2013 {\em Phys.\ Rev.\ D\/} {\bf 88} 043007

\bibitem{LIGOIndia}
Iyer B, Souradeep T, Unnikrishnan C, Dhurandhar S, Raja S and Sengupta A 2012
  Ligo-india, proposal of the consortium for indian initiative in
  gravitational-wave observations (indigo) Tech. Rep. LIGO-M1100296
  \url{https://dcc.ligo.org/LIGO-M1100296/public}

\bibitem{EinsteinTelescope}
{Punturo} M, {Abernathy} M, {Acernese} F, {Allen} B, {Andersson} N, {Arun} K,
  {Barone} F, {Barr} B, {Barsuglia} M, {Beker} M, {Beveridge} N, {Birindelli}
  S, {Bose} S, {Bosi} L, {Braccini} S, {Bradaschia} C, {Bulik} T, {Calloni} E,
  {Cella} G, {Chassande Mottin} E, {Chelkowski} S, {Chincarini} A, {Clark} J,
  {Coccia} E, {Colacino} C, {Colas} J, {Cumming} A, {Cunningham} L, {Cuoco} E,
  {Danilishin} S, {Danzmann} K, {De Luca} G, {De Salvo} R, {Dent} T, {Derosa}
  R, {Di Fiore} L, {Di Virgilio} A, {Doets} M, {Fafone} V, {Falferi} P,
  {Flaminio} R, {Franc} J, {Frasconi} F, {Freise} A, {Fulda} P, {Gair} J,
  {Gemme} G, {Gennai} A, {Giazotto} A, {Glampedakis} K, {Granata} M, {Grote} H,
  {Guidi} G, {Hammond} G, {Hannam} M, {Harms} J, {Heinert} D, {Hendry} M,
  {Heng} I, {Hennes} E, {Hild} S, {Hough} J, {Husa} S, {Huttner} S, {Jones} G,
  {Khalili} F, {Kokeyama} K, {Kokkotas} K, {Krishnan} B, {Lorenzini} M,
  {L{\"u}ck} H, {Majorana} E, {Mandel} I, {Mandic} V, {Martin} I, {Michel} C,
  {Minenkov} Y, {Morgado} N, {Mosca} S, {Mours} B, {M{\"u}ller-Ebhardt} H,
  {Murray} P, {Nawrodt} R, {Nelson} J, {Oshaughnessy} R, {Ott} C~D, {Palomba}
  C, {Paoli} A, {Parguez} G, {Pasqualetti} A, {Passaquieti} R, {Passuello} D,
  {Pinard} L, {Poggiani} R, {Popolizio} P, {Prato} M, {Puppo} P, {Rabeling} D,
  {Rapagnani} P, {Read} J, {Regimbau} T, {Rehbein} H, {Reid} S, {Rezzolla} L,
  {Ricci} F, {Richard} F, {Rocchi} A, {Rowan} S, {R{\"u}diger} A, {Sassolas} B,
  {Sathyaprakash} B, {Schnabel} R, {Schwarz} C, {Seidel} P, {Sintes} A,
  {Somiya} K, {Speirits} F, {Strain} K, {Strigin} S, {Sutton} P, {Tarabrin} S,
  {van den Brand} J, {van Leewen} C, {van Veggel} M, {van den Broeck} C,
  {Vecchio} A, {Veitch} J, {Vetrano} F, {Vicere} A, {Vyatchanin} S, {Willke} B,
  {Woan} G, {Wolfango} P and {Yamamoto} K 2010 {\em Class.\ Quantum Grav.\/}
  {\bf 27} 084007

\bibitem{abbott2017exploring}
{{B~P~Abbott et al~for the LIGO Scientific Collaboration}} 2017 {\em Classical
  and Quantum Gravity\/} {\bf 34} 044001

\bibitem{aLIGO2}
Aasi J {\em et~al.\/} (LIGO Scientific Collaboration) 2015 {\em Class.\ Quantum
  Grav.\/} {\bf 32} 074001 (\textit{Preprint} \eprint{1411.4547})

\bibitem{gwinc}
https://awiki.ligo-wa.caltech.edu/aLIGO/GWINC

\bibitem{adhikari2014gravitational}
Adhikari R~X 2014 {\em Reviews of modern physics\/} {\bf 86} 121

\bibitem{harry2002thermal}
Harry G~M, Gretarsson A~M, Saulson P~R, Kittelberger S~E, Penn S~D, Startin
  W~J, Rowan S, Fejer M~M, Crooks D, Cagnoli G, Hough J and Nakagawa N 2002
  {\em Class.\ Quantum Grav.\/} {\bf 19} 897

\bibitem{penn2003mechanical}
Penn S~D, Sneddon P~H, Armandula H, Betzwieser J~C, Cagnoli G, Camp J, Crooks
  D, Fejer M~M, Gretarsson A~M, Harry G~M, Hough J, Kittelberger S~E, Mortonson
  M~J, Route R, Rowan S and Vassiliou C~C 2003 {\em Class.\ Quantum Grav.\/}
  {\bf 20} 2917

\bibitem{rowan2005thermal}
Rowan S, Hough J and Crooks D 2005 {\em Physics Letters A\/} {\bf 347} 25--32

\bibitem{harry2006thermal}
Harry G~M, Armandula H, Black E, Crooks D, Cagnoli G, Hough J, Murray P, Reid
  S, Rowan S, Sneddon P, Fejer M~M, Route R and Penn S~D 2006 {\em Applied
  optics\/} {\bf 45} 1569--1574

\bibitem{harry2007titania}
Harry G~M, Abernathy M~R, Becerra-Toledo A~E, Armandula H, Black E, Dooley K,
  Eichenfield M, Nwabugwu C, Villar A, Crooks D, Cagnoli G, Hough J, How C~R,
  MacLaren I, Murray P, Reid S, Rowan S, Sneddon P~H, Fejer M~M, Route R, Penn
  S~D, Ganau P, Mackowski J~M, Michel C, Pinard L and Remillieux A 2007 {\em
  Class.\ Quantum Grav.\/} {\bf 24} 405

\bibitem{flaminio2010study}
Flaminio R, Franc J, Michel C, Morgado N, Pinard L and Sassolas B 2010 {\em
  Class.\ Quantum Grav.\/} {\bf 27} 084030

\bibitem{kondratiev2011thermal}
Kondratiev N, Gurkovsky A and Gorodetsky M 2011 {\em Phys.\ Rev.\ D\/} {\bf 84}
  022001

\bibitem{evans2012reduced}
Evans K, Bassiri R, Maclaren I, Rowan S, Martin I, Hough J and Borisenko K 2012
  Reduced density function analysis of titanium dioxide doped tantalum
  pentoxide {\em Journal of Physics: Conference Series\/} vol 371 (IOP
  Publishing) p 012058

\bibitem{hong2013brownian}
Hong T, Yang H, Gustafson E~K, Adhikari R~X and Chen Y 2013 {\em Phys.\ Rev.\
  D\/} {\bf 87} 082001

\bibitem{bassiri2013correlations}
Bassiri R, Evans K, Borisenko K, Fejer M, Hough J, MacLaren I, Martin I, Route
  R and Rowan S 2013 {\em Acta Materialia\/} {\bf 61} 1070--1077

\bibitem{murray2017cryogenic}
Murray P~G, Martin I~W, Craig K, Hough J, Rowan S, Bassiri R, Fejer M~M, Harris
  J~S, Lantz B~T, Lin A~C {\em et~al.\/} 2017 {\em Physical Review D\/} {\bf
  95} 042004

\bibitem{gras2017audio}
Gras S, Yu H, Yam W, Martynov D and Evans M 2017 {\em Physical Review D\/} {\bf
  95} 022001

\bibitem{kroker2017brownian}
Kroker S, Dickmann J, Hurtado C, Heinert D, Nawrodt R, Levin Y and Vyatchanin S
  2017 {\em arXiv preprint arXiv:1705.00157\/}

\bibitem{braginsky2003thermodynamical}
Braginsky V and Vyatchanin S 2003 {\em Physics Letters A\/} {\bf 312} 244--255

\bibitem{fejer2004thermoelastic}
Fejer M, Rowan S, Cagnoli G, Crooks D, Gretarsson A, Harry G, Hough J, Penn S,
  Sneddon P and Vyatchanin S 2004 {\em Phys.\ Rev.\ D\/} {\bf 70} 082003

\bibitem{evans2008thermo}
Evans M, Ballmer S, Fejer M, Fritschel P, Harry G and Ogin G 2008 {\em Phys.\
  Rev.\ D\/} {\bf 78} 102003

\bibitem{braginsky2000thermo}
Braginsky V, Gorodetsky M and Vyatchanin S 2000 {\em Physics Letters A\/} {\bf
  271} 303--307

\bibitem{liu2000thermoelastic}
Liu Y~T and Thorne K~S 2000 {\em Phys.\ Rev.\ D\/} {\bf 62} 122002

\bibitem{o2004implications}
O'Shaughnessy R, Strigin S and Vyatchanin S 2004 {\em arXiv preprint
  gr-qc/0409050\/}

\bibitem{vinet2005mirror}
Vinet J~Y 2005 {\em Class.\ Quantum Grav.\/} {\bf 22} 1395

\bibitem{saulson1990thermal}
Saulson P~R 1990 {\em Phys.\ Rev.\ D\/} {\bf 42} 2437

\bibitem{gillespie1993thermal}
Gillespie A and Raab F 1993 {\em Physics Letters A\/} {\bf 178} 357--363

\bibitem{braginsky1999reduce}
Braginsky V~B, Levin Y and Vyatchanin S 1999 {\em Measurement Science and
  Technology\/} {\bf 10} 598

\bibitem{gonzalez2000suspensions}
Gonz{\'a}lez G 2000 {\em Class.\ Quantum Grav.\/} {\bf 17} 4409

\bibitem{kumar2008finite}
Kumar R 2008 {\em Finite element analysis of suspension elements for
  gravitational wave detectors\/} Ph.D. thesis University of Glasgow

\bibitem{Harry:2011book}
Harry G, Bodiya T~P and DeSalvo R (eds) 2011 {\em Optical Coatings and Thermal
  Noise in Precision Measurement\/} (Cambridge University Press)

\bibitem{chalermsongsak2015coherent}
Chalermsongsak T, Hall E~D, Cole G~D, Follman D, Seifert F, Arai K, Gustafson
  E~K, Smith J~R, Aspelmeyer M and Adhikari R~X 2015 {\em arXiv preprint
  arXiv:1506.07088\/}

\bibitem{jiang2011making}
Jiang Y, Ludlow A, Lemke N, Fox R, Sherman J, Ma L~S and Oates C 2011 {\em
  Nature Photonics\/} {\bf 5} 158--161

\bibitem{harry2012optical}
Harry G, Bodiya T~P and DeSalvo R (eds) 2012 {\em Optical coatings and thermal
  noise in precision measurement\/} (Cambridge, United Kingdom: Cambridge
  University Press)

\bibitem{callen1951irreversibility}
Callen H~B and Welton T~A 1951 {\em Physical Review\/} {\bf 83} 34

\bibitem{bernard1959fdt}
Bernard W and Callen H~B 1959 {\em Rev.~Mod.~Phys.\/} {\bf 31} 1017

\bibitem{kubo1966fluctuation}
Kubo R 1966 {\em Reports on progress in physics\/} {\bf 29} 255

\bibitem{gonzalez1994brownian}
Gonz{\'a}lez G~I and Saulson P~R 1994 {\em The Journal of the Acoustical
  Society of America\/} {\bf 96} 207--212

\bibitem{levin1998internal}
Levin Y 1998 {\em Phys.\ Rev.\ D\/} {\bf 57} 659

\bibitem{hirose2014update}
Hirose E, Sekiguchi T, Kumar R and {(for the KAGRA collaboration)} R~T 2014
  {\em Class.\ Quantum Grav.\/} {\bf 31} 224004

\bibitem{punturo2010einstein}
Punturo M, Abernathy M, Acernese F, Allen B, Andersson N, Arun K, Barone F,
  Barr B, Barsuglia M, Beker M, Beveridge N, Birindelli S, Bose S, Bosi L,
  Braccini S, Bradaschia C, Bulik T, Calloni E, Cella G, Mottin E~C, Chelkowski
  S, Chincarini A, Clark J, Coccia E, Colacino C, Colas J, Cumming A,
  Cunningham L, Cuoco E, Danilishin S, Danzmann K, Luca G~D, Salvo R~D, Dent T,
  Rosa R~D, Fiore L~D, Virgilio A~D, Doets M, Fafone V, Falferi P, Flaminio R,
  Franc J, Frasconi F, Freise A, Fulda P, Gair J, Gemme G, Gennai A, Giazotto
  A, Glampedakis K, Granata M, Grote H, Guidi G, Hammond G, Hannam M, Harms J,
  Heinert D, Hendry M, Heng I, Hennes E, Hild S, Hough J, Husa S, Huttner S,
  Jones G, Khalili F, Kokeyama K, Kokkotas K, Krishnan B, Lorenzini M, L\"{u}ck
  H, Majorana E, Mandel I, Mandic V, Martin I, Michel C, Minenkov Y, Morgado N,
  Mosca S, Mours B, M\"{u}ller-Ebhardt H, Murray P, Nawrodt R, Nelson J,
  Oshaughnessy R, Ott C~D, Palomba C, Paoli A, Parguez G, Pasqualetti A,
  Passaquieti R, Passuello D, Pinard L, Poggiani R, Popolizio P, Prato M, Puppo
  P, Rabeling D, Rapagnani P, Read J, Regimbau T, Rehbein H, Reid S, Rezzolla
  L, Ricci F, Richard F, Rocchi A, Rowan S, R\"{u}diger A, Sassolas B,
  Sathyaprakash B, Schnabel R, Schwarz C, Seidel P, Sintes A, Somiya K,
  Speirits F, Strain K, Strigin S, Sutton P, Tarabrin S, Th\"{u}ring A, van~den
  Brand J, van Leewen C, van Veggel M, van~den Broeck C, Vecchio A, Veitch J,
  Vetrano F, Vicere A, Vyatchanin S, Willke B, Woan G, Wolfango P and Yamamoto
  K 2010 {\em Class.\ Quantum Grav.\/} {\bf 27} 194002

\bibitem{schroeter2007mechanical}
Schroeter A, Nawrodt R, Schnabel R, Reid S, Martin I, Rowan S, Schwarz C,
  Koettig T, Neubert R, Th{\"u}rk M, Vodel W, T\'{u}nnermann A, Danzmann K and
  Seidel P 2007 {\em arXiv preprint arXiv:0709.4359\/}

\bibitem{cole2013tenfold}
Cole G~D, Zhang W, Martin M~J, Ye J and Aspelmeyer M 2013 {\em Nature
  Photonics\/} {\bf 7} 644--650

\bibitem{cole2016highperformance}
Cole G~D, Zhang W, Bjork B~J, Follman D, Heu P, Deutsch C, Sonderhouse L,
  Robinson J, Frans C, Alexandrovski A, Notcutt M, Heckl O~H, Ye J and
  Aspelmeyer M 2016 {\em Optica\/} {\bf 3} 647--656 (\textit{Preprint}
  \eprint{1604.00065})

\bibitem{pang2009effect}
Pang Y and Huang R 2009 {\em International Journal of Solids and Structures\/}
  {\bf 46} 2822--2833

\bibitem{heinert2014fluctuation}
Heinert D, Hofmann G and Nawrodt R 2014 Fluctuation dissipation at work:
  Thermal noise in reflective optical coatings for gw detectors {\em Metrology
  for Aerospace (MetroAeroSpace), 2014 IEEE\/} (IEEE) pp 293--298

\bibitem{abernathyMultipleLossAngles}
Abernathy M, Harry G, Newport J, Fair H, Kinley-Hanlon M, Hickey S, Jiffar I,
  Thompson C, Gretarsson A, Penn S, Basirie R, Gustafson E and Martin I Bulk
  and shear mechanical loss of titania-doped tantala Tech. Rep. LIGO-P1700093

\bibitem{ibrahimbegovic2009nonlinear}
Ibrahimbegovic A 2009 {\em Nonlinear solid mechanics: theoretical formulations
  and finite element solution methods\/} vol 160 (Springer Science \& Business
  Media)

\bibitem{step8}
\url{https://www.dealii.org/8.4.0/doxygen/deal.II/step_8.html}

\bibitem{dealII82}
Bangerth W, Heister T, Heltai L, Kanschat G, Kronbichler M, Maier M, Turcksin B
  and Young T 2015 {\em Archive of Numerical Software\/} {\bf 3} ISSN 2197-8263
  \urlprefix\url{http://journals.ub.uni-heidelberg.de/index.php/ans/article/view/18031}

\bibitem{dealII85}
Arndt D, Bangerth W, Davydov D, Heister T, Heltai L, Kronbichler M, Maier M,
  Pelteret J~P, Turcksin B and Wells D 2017 {\em Journal of Numerical
  Mathematics\/}

\bibitem{petsc-web-page}
Balay S, Buschelman K, Gropp W~D, Kaushik D, Knepley M~G, McInnes L~C, Smith
  B~F and Zhang H 2009 {PETSc} {W}eb {P}age \url{http://www.mcs.anl.gov/petsc}

\bibitem{petsc-user-ref}
Balay S, Buschelman K, Eijkhout V, Gropp W~D, Kaushik D, Knepley M~G, McInnes
  L~C, Smith B~F and Zhang H 2008 {PETS}c users manual Tech. Rep. ANL-95/11 -
  Revision 3.0.0 Argonne National Laboratory

\bibitem{petsc_efficient}
Balay S, Gropp W~D, McInnes L~C and Smith B~F 1997 Efficient management of
  parallelism in object oriented numerical software libraries {\em Modern
  Software Tools in Scientific Computing\/} ed Arge E, Bruaset A~M and
  Langtangen H~P (Boston: Birkh{\"a}user Press) pp 163--202

\bibitem{falgout2002hypre}
Falgout R and Yang U 2002 {\em Computational Science—ICCS 2002\/}  632--641

\bibitem{kelly1983posteriori}
Kelly D, Gago D~S, Zienkiewicz O and Babuska I 1983 {\em International journal
  for numerical methods in engineering\/} {\bf 19} 1593--1619

\bibitem{gago1983posteriori}
Gago D~S, Kelly D, Zienkiewicz O and Babuska I 1983 {\em International journal
  for numerical methods in engineering\/} {\bf 19} 1621--1656

\bibitem{gehrsitz1999compositional}
Gehrsitz S, Sigg H, Herres N, Bachem K, K{\"o}hler K and Reinhart F 1999 {\em
  Physical Review B\/} {\bf 60} 11601

\bibitem{thermalNoiseRepo}
Lovelace G, Demos N and Khan H 2017
  \url{https://git.ligo.org/geoffrey-lovelace/NumericalCoatingThermalNoise}

\bibitem{heptonstall2014}
Heptonstall A, Barton M~A, Bell A~S, Bohn A, Cagnoli G, Cumming A, Grant A,
  Gustafson E, Hammond G~D, Hough J, Jones R, Kumar R, Lee K, Martin I~W,
  Robertson N~A, Rowan S, Strain K~A and Tokmakov K~V 2014 {\em Classical and
  Quantum Gravity\/} {\bf 31} 105006
  \urlprefix\url{http://stacks.iop.org/0264-9381/31/i=10/a=105006}

\bibitem{o2006note}
O'Shaughnessy R 2006 {\em Class.\ Quantum Grav.\/} {\bf 23} 7627

\bibitem{Lovelace:2006ne}
Lovelace G 2007 {\em Class.\ Quantum Grav.\/} {\bf 24} 4491--4512
  (\textit{Preprint} \eprint{gr-qc/0610041})

\end{thebibliography}

\end{document}